\documentclass[aps,prl,twocolumn,superscriptaddress,showpacs]{revtex4-2}
\usepackage{amsmath,amssymb,wasysym,graphicx}
\usepackage[utf8]{inputenc}
\usepackage[T1]{fontenc}
\usepackage{xcolor}

\IfFileExists{newtxtext.sty}
{\usepackage{newtxtext,newtxmath}}
{\IfFileExists{stix.sty}
	{\usepackage{stix}}
	{\IfFileExists{mathptmx.sty}
		{\usepackage{mathptmx}}{} } }

\usepackage{textcomp}

\usepackage{bm}

\IfFileExists{siunitx.sty}{\usepackage{booktabs,siunitx}}{}

\usepackage{color}
\definecolor{LinkColor}{rgb}{0.256,0.439,0.588}
\usepackage{hyperref}
\hypersetup{
	pdfauthor={good guys},
	pdftitle={good title},
	colorlinks=true,
	citecolor=LinkColor,
	linkcolor=LinkColor,
	urlcolor=LinkColor
}

\def\normOrd#1{\mathop{:}\nolimits\!#1\!\mathop{:}\nolimits}
\newcommand{\beq} {\begin{equation}}
\newcommand{\eeq} {\end{equation}}
\newcommand{\bea} {\begin{eqnarray}}
\newcommand{\eea} {\end{eqnarray}}
\newcommand{\be} {\begin{equation}}
\newcommand{\ee} {\end{equation}}
\renewcommand{\(}{\left(}
\renewcommand{\)}{\right)}
\renewcommand{\[}{\left[}
\renewcommand{\]}{\right]}

\newcommand{\ket}[1]{\left|#1\right>}
\newcommand{\bra}[1]{\left<#1\right|}

\def\Eq#1{Eq.~(\ref{#1})}
\def\Fig#1{Fig.~\ref{#1}}
\def\avg#1{\left\langle#1\right\rangle}

\begin{document}

\title{Gifts from long-range interaction: Emergent gapless topological behaviors in quantum spin chain}

\author{Sheng Yang}
\affiliation{Institute for Advanced Study in Physics and School of Physics, Zhejiang University, Hangzhou 310058, China}

\author{Hai-Qing Lin}
\email{hqlin@zju.edu.cn}
\affiliation{Institute for Advanced Study in Physics and School of Physics, Zhejiang University, Hangzhou 310058, China}

\author{Xue-Jia Yu}
\email{xuejiayu@fzu.edu.cn}
\affiliation{Department of Physics, Fuzhou University, Fuzhou 350116, Fujian, China}
\affiliation{Fujian Key Laboratory of Quantum Information and Quantum Optics,
College of Physics and Information Engineering,
Fuzhou University, Fuzhou, Fujian 350108, China}

\begin{abstract}
Topology in condensed matter physics is typically associated with a bulk energy gap. However, recent research has shifted focus to topological phases without a bulk energy gap, exhibiting nontrivial gapless topological behaviors. In this letter, we explore a cluster Ising chain with long-range antiferromagnetic interactions that decay as a power law with the distance. Using complementary numerical and analytical techniques, we demonstrate that long-range interactions can unambiguously induce an algebraic topological phase and a topological Gaussian universality, both of which exhibit nontrivial gapless topological behaviors. Our study not only provides a platform to investigate the fundamental physics of quantum many-body systems but also offers a novel route toward searching for gapless topological phases in realistic quantum simulators.

\end{abstract}

\maketitle

\emph{Introduction.}---Locality is a fundamental concept in quantum many-body physics, implying that the Hamiltonian consists of a sum of terms with local interactions~\cite{sachdev2023quantum,sachdev_2011,fradkin2013field,cardy1996scaling}. However, describing many-body systems in terms of local interactions is often an approximation and not always a reliable one~\cite{defenu2023rmp}. Long-range (LR) power-law interactions ($1/r^{\alpha}$) are typical non-local interactions that are both of fundamental and practical importance, appearing ubiquitously in nature and in numerous experimental systems of current interest, such as trapped-ion~\cite{kyprianidis2021observation,pagano2020quantum}, nitrogen-vacancy centers~\cite{schirhagl2014nitrogen,Ho2018prl}, and programmable Rydberg atom arrays~\cite{labuhn2016tunable,saffman2010rmp}. On the theoretical side, LR interactions are widespread and lead to qualitatively new physics, including modifications to the Mermin-Wagner~\cite{bruno2001prl,maghrebi2017prl} and Lieb-Schultz-Mattis theorems~\cite{liu2024liebschultzmattis,ma2024liebschultzmattis}, the breakdown of the entanglement area law for gapped phases or the conformal symmetry at criticality~\cite{vodola2014prl,Vodola_2016,Ares2018pra,Koffel2012prl}, the emergence of new Lieb-Robinson bounds~\cite{Hauke2013prl,gong2023prl}, as well as modifying the critical behavior at phase transitions~\cite{dutta2001prb,Laflorencie_2005,Defenu2017prb,fey2019prl,yu2022prb,yu2023pre,yu2023prb,xiao2024twodimensional}.

On a different front, symmetry-protected topological (SPT) phases, as a typical class of quantum matter beyond the Landau paradigm, have attracted much attention in the past few decades~\cite{gu2009prb,chen2013prb,wen2017rmp,senthil2015symmetry,pollmann2010prb}. Although the bulk gap plays a crucial role in defining SPT phases, recent progress~\cite{keselman2015prb,meng2011prb,fidkowski2011prb,kestner2011prb,Iemini2015prl,Lang2015prb,scaffidi2017prx,ruhman2017prb,JIANG2018753,verresen2018prl,keselman2018prb,verresen2020topology,verresen2021prx,duque2021prb,thorngren2021prb,umberto2021sci_post,yu2022prl,parker2018prb,yu2024universal,yu2024quantum,zhong2024topological,li2023intrinsicallypurely,li2023decorated,huang2023topological,wen2023prb,wen2023classification,su2024gapless,zhang2024quantum} has revealed that many key features of topological properties persist even with nontrivial coupling between the boundary and the gapless bulk. This extension is termed \emph{gapless topological phases}~\cite{scaffidi2017prx,verresen2018prl,verresen2021prx,verresen2020topology}, which exhibit nontrivial gapless topological behaviors, including symmetry-protected degenerate edge modes, power-law decay in string order parameters~\cite{verresen2021prx}, nontrivial conformal boundary conditions~\cite{yu2022prl,parker2018prb}, and universal bulk-boundary correspondence in the entanglement spectrum~\cite{yu2024universal}.

\begin{figure}[tb]
    \includegraphics[width=0.9\linewidth]{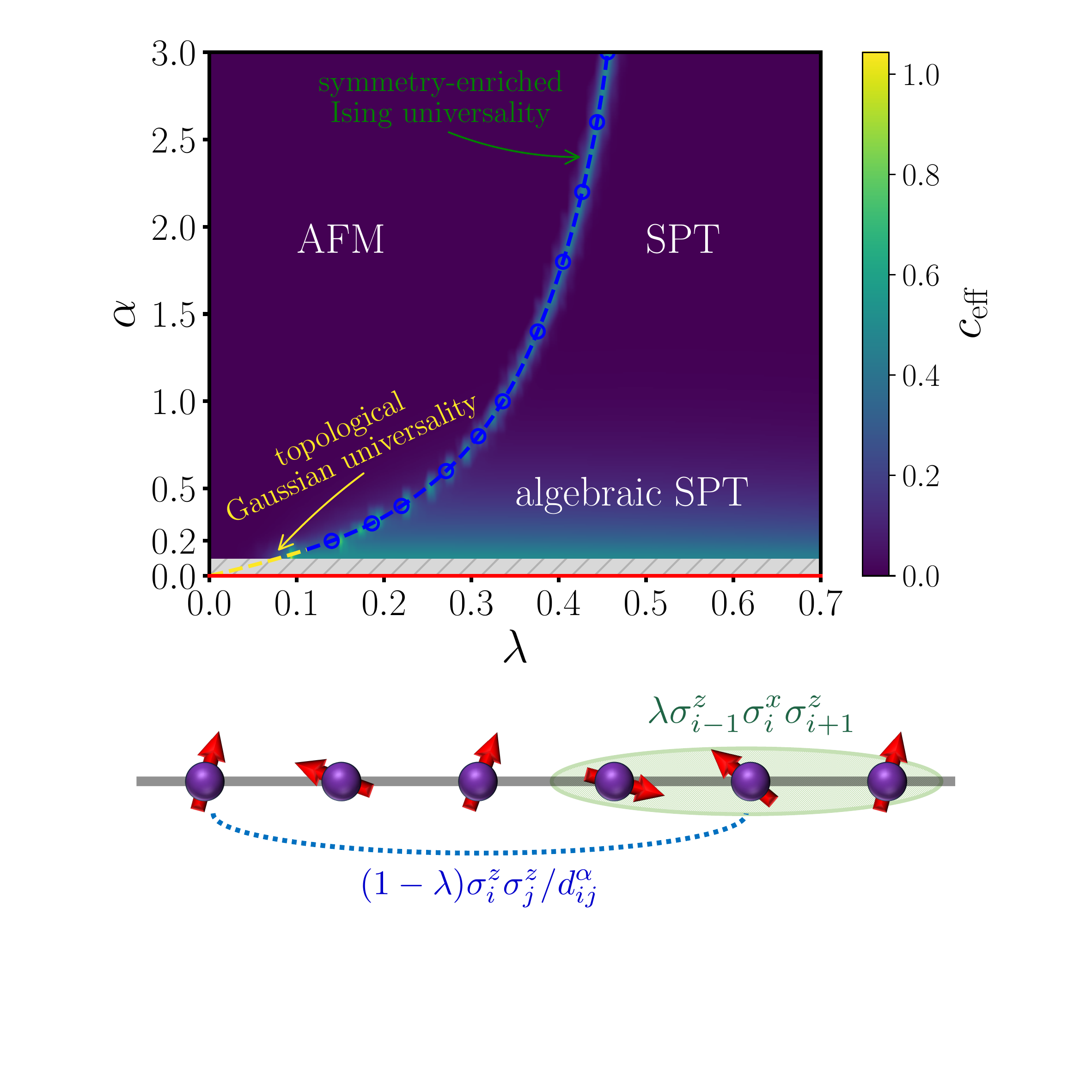}
    \caption{Summary of the main results. Below the phase diagram is a schematic of the LR cluster Ising chain, where the blue dashed line (green-filled ellipsoid) represents the LR AFM interaction (cluster interaction). The ground-state phase diagram of the Hamiltonian~\eqref{E1} exhibits AFM, SPT, and algebraic SPT phases, along with symmetry-enriched Ising and topological Gaussian (yellow dashed line) universality classes. The red line represents a new gapped SPT phase with two-fold degeneracy. The blue markers are determined by the Binder ratio crossings, and the colored background indicates the effective central charge $c_{\text{eff}}$ defined in the main text. The gray region ($\alpha < 0.1$) is beyond our computation ability.}
    \label{fig:phasediagram}
\end{figure}

From an experimental point of view, it is inevitable to consider the influence of LR interactions when implementing nontrivial topological phases in state-of-the-art quantum simulators. To date, various studies~\cite{gong2016prb,Maity_2020,patrick2017prl} have demonstrated qualitative changes of physical properties under LR interactions in gapped topological phases, including the emergence of new massive edge modes~\cite{vodola2014prl,viyuela2016prb,Vodola_2016} and novel bulk-boundary correspondence~\cite{Lepori_2017,jones2023prl}. However, it is not immediately clear whether LR interactions can have a significant impact on gapless topological phases. This is because the gapless bulk is usually sensitive to LR interactions, while topological properties remain robust against them. Therefore, it is valuable to address the following intriguing questions: how does the gapless topological phase respond to LR interactions, and do LR interactions induce new phases or phase transitions with nontrivial gapless topological behaviors? If so, how to understand the underlying mechanism?

To conclusively address the aforementioned issues, in this letter, we focus on investigating the simplest type of gapless topological phase---symmetry-enriched or topological nontrivial quantum critical points (QCPs). Notably, we concentrate on a one-dimensional (1D) cluster Ising model featuring LR antiferromagnetic (AFM) interactions, which is amenable to powerful density matrix renormalization group (DMRG) simulations~\cite{white1992prl,white1993prb,SCHOLLWOCK201196,schollwock2005rmp,Cirac2006PRB}. We conduct comprehensive simulations to investigate the ground-state phase diagram as a function of the cluster interaction strength $\lambda$ and the LR power exponent $\alpha$. For large $\alpha \textgreater 1.0$, the ground state manifests as an AFM (cluster SPT) phase for $\lambda$ below (above) a critical value $\lambda_{\rm c}$, separating a symmetry-enriched Ising critical point. Conversely, for sufficiently small $\alpha \lesssim 1.0$, the cluster SPT phase gradually crossovers to a different SPT phase characterized by algebraic correlations. In the limit $\alpha=0$, the ground state is a gapped topological phase with two-fold degenerate edge modes for any $\lambda \textgreater 0$. Meanwhile, the universality class of the phase transition gradually changes from topological Ising to the topological Gaussian universality as $\alpha$ decreases to $0$. Finally, we provide an intuitive physical picture to elucidate the numerical observations.

\emph{Model and method.}---We consider the following LR interacting Hamiltonian for spins on a lattice of length $L$ (see Fig.~\ref{fig:phasediagram})
\begin{equation}
    \begin{split}
    \label{E1}
    H_{\text{LRCI}} = \sum_{i<j}\frac{(1-\lambda)}{d_{ij}^{\alpha}}\sigma^{z}_{i}\sigma^{z}_{j}+\lambda \sum_{j}\sigma^{z}_{j-1}\sigma^{x}_{j}\sigma^{z}_{j+1} \, ,
    \end{split}
\end{equation}
where $\vec{\sigma_{i}}=(\sigma^{x}_{i},\sigma^{y}_{i},\sigma^{z}_{i})$ represents the Pauli matrix on site $i$. The parameter $\alpha$ is the power exponent of the LR interaction, and $d_{ij}$ is the distance between sites $i$ and $j$, which is $|i-j|$ and $\min(|i-j|, L-|i-j|)$ for open (OBC) and periodic (PBC) boundary conditions, respectively. It is noted that the LR AFM interactions compete with each other and the ground-state energy is still an extensive quantity linear with $L$, therefore, the implementation of a Kac rescaling is not necessary in our case (see Sec.~I of Supplementary Materials (SM)). Notably, for $\alpha \to \infty$, Eq.~\eqref{E1} reduces to the nearest neighbor short-range (SR) cluster Ising model, which can be exactly solved via the Jordan-Wigner transformation~\cite{Son_2011,verresen2017prb,smacchia2011pra,ding2019pre,guo2022pra}. The parameter $\lambda$ governs the competition between the Ising and cluster interactions, driving the system toward different phases, including AFM and cluster SPT phases~\cite{Son_2011}. The phase transition between them belongs to the symmetry-enriched Ising universality class, featuring topological-protected edge modes with two-fold degeneracy~\cite{verresen2018prl,verresen2021prx,yu2024quantum,yu2024universal}. Conversely, for $\alpha=0$, the all-to-all interacting model can be solved via the Holstein-Primakoff transformation, leading to a gapped topological phase with two-fold degenerate edge modes for any $\lambda \textgreater 0$ (see Sec.~VII of SM for details).

For general $\alpha$, Eq.~\eqref{E1} is no longer exactly solvable and we can only resort to numerical simulations. In this work, we employ a large-scale DMRG simulation to study the global ground-state phase diagram of the LR model. The numerical details are introduced in Sec.~I of the SM. We focus on the model for both sufficiently strong and weak LR interactions, exploring the exotic phases and phase transitions arising from the interplay between topology and LR interactions.

\begin{figure}[tb]
    \includegraphics[width=0.9\linewidth]{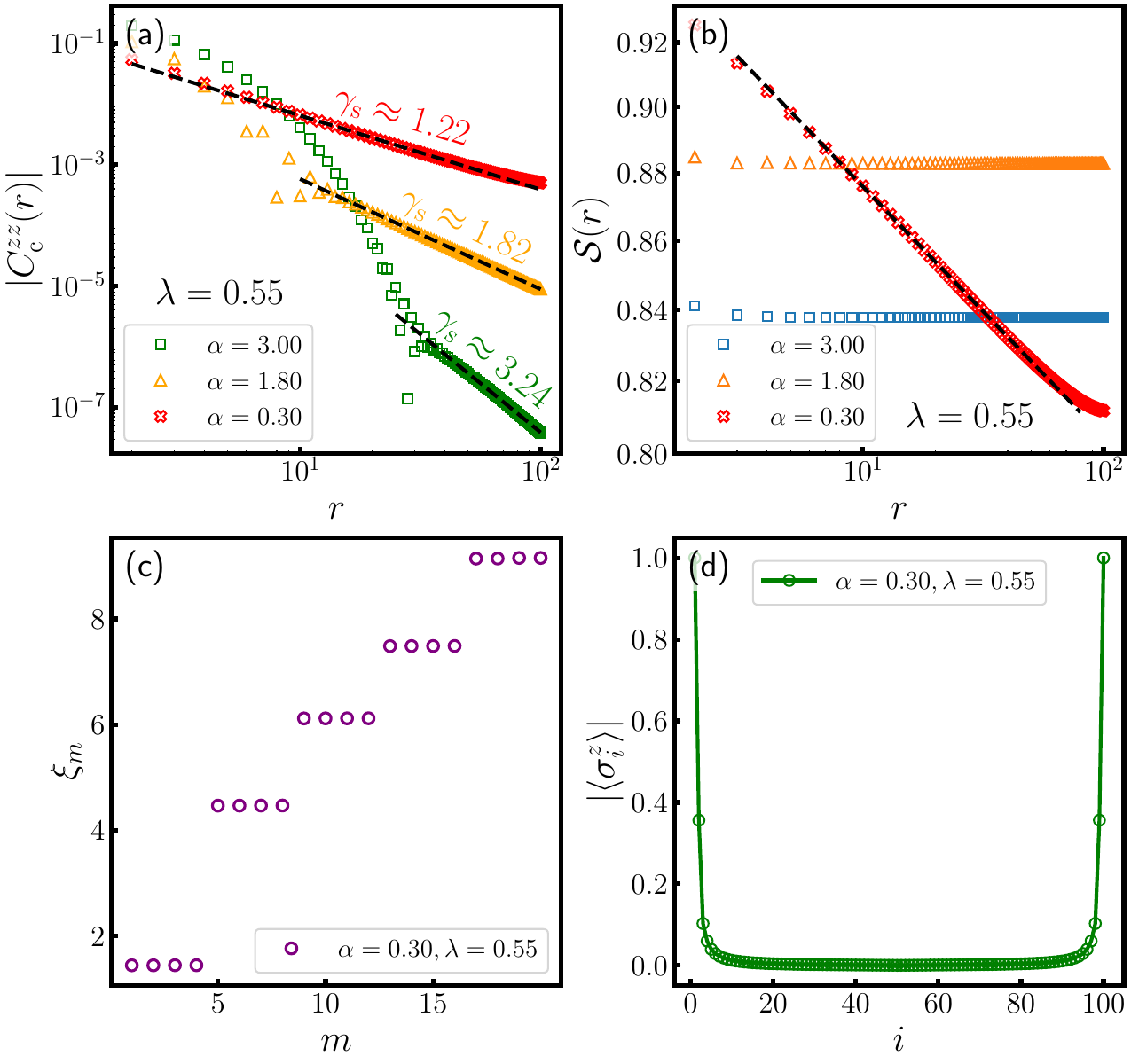}
    \caption{The double logarithmic plot of the connected spin-spin correlation $|C_{\rm c}^{zz}(r)|$ (a) and the SPT string order parameter $\mathcal{S}(r)$ (b) versus the lattice distance $r$ for $\alpha = 0.3, 1.8, 3.0$ with fixed $\lambda=0.55$\,. The bulk entanglement spectrum (c) and magnetization profile (d) within the typical algebraic SPT region for $\alpha=0.3$ and $\lambda=0.55$\,. $m$ counts the spectrum from the lowest-lying levels and $i$ labels the site position.}
    \label{fig2}
\end{figure}

\begin{figure}[tb]
    \includegraphics[width=0.9\linewidth]{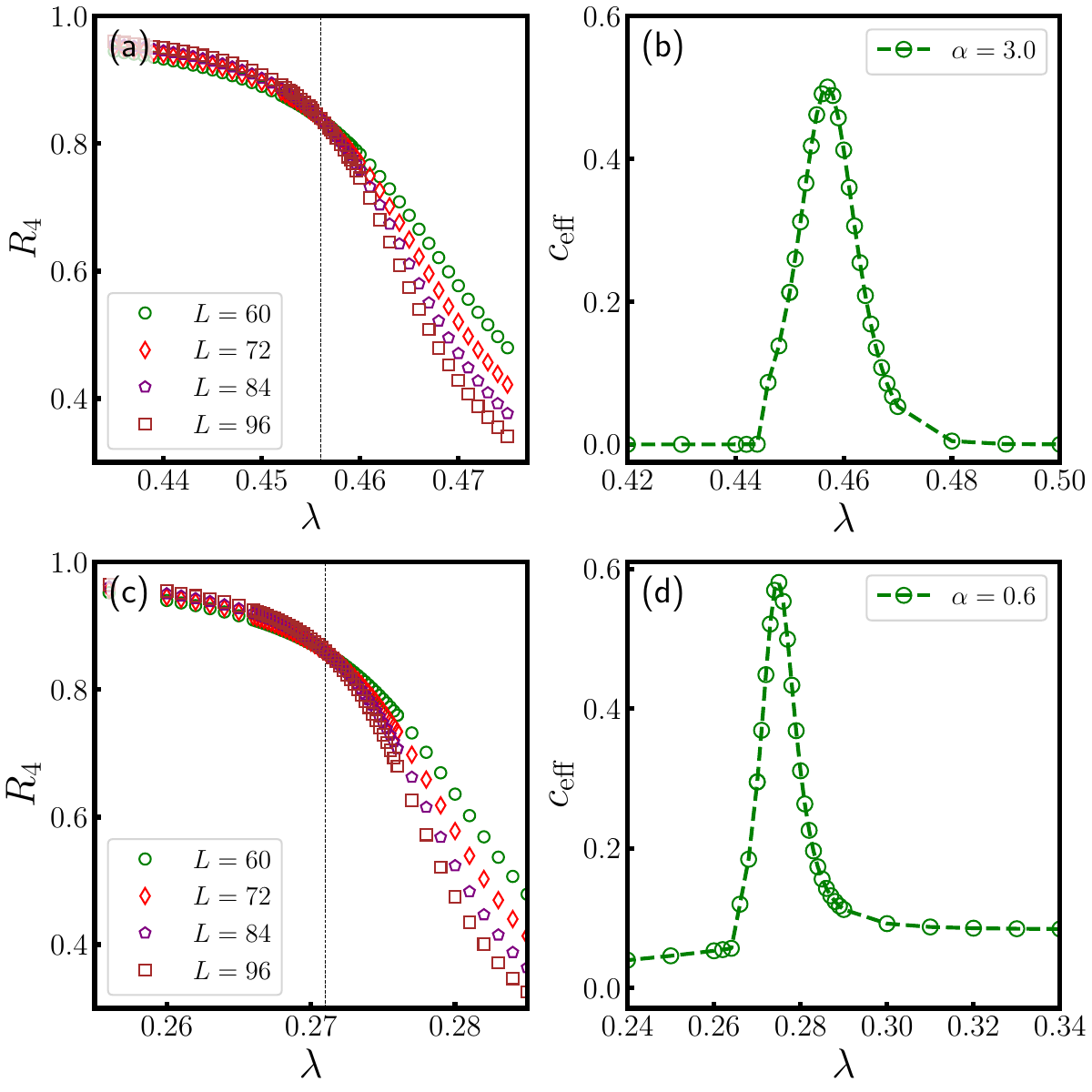}
    \caption{The Binder ratio $R_{4}$ for AFM order versus cluster interaction strength $\lambda$ for $\alpha=3.0$ (a) and $0.6$ (c). The effective central charge as a function of $\lambda$ for $\alpha = 3.0$ (b) and $0.6$ (d). The simulated system size is $L=96$ under PBC for (b) and (d).}
    \label{fig3}
\end{figure}

\emph{Quantum phase diagram.}---Before presenting the details of the DMRG results, we first summarize our main findings. A schematic phase diagram with varying LR power $\alpha$ and cluster interaction strength $\lambda$ is shown in Fig.~\ref{fig:phasediagram}. For large $\alpha \textgreater 1.0 $, the ground state can exhibit AFM and cluster SPT phases separated by a symmetry-enriched Ising critical point. As $\alpha$ decreases, the LR quantum fluctuation is enhanced, resulting in exotic phases and phase transitions. In the limit $\alpha = 0$, the LR interaction becomes all-to-all interactions. We can solve the model via the Holstein-Primakoff transformation and show that the ground state is a gapped topological phase exhibiting two-fold degenerate edge modes for any $\lambda \textgreater 0$ (the red line in Fig.~\ref{fig:phasediagram}). For sufficiently small $\alpha \lesssim 1.0$, we find that the connected spin-spin correlation (string order parameter) changes from a hybrid behavior---exponential at short distances and algebraic (constant) at long distances---to purely algebraic decay for $\lambda > \lambda_{\rm c}$. This suggests that the gapped SPT phase gradually crossovers to a new algebraic SPT phase (define below). Notably, the entire critical line exhibits robust two-fold degenerate edge modes, and the universality class gradually changes from topological Ising with central charge $c=1/2$ to the topological Gaussian universality class with effective central charge $c_{\rm eff}=1$ as $\alpha$ approaches $0$, indicating that LR interactions can induce a new topologically nontrivial universality class.

\emph{Algebraic SPT phase.}---To investigate the effect of LR interactions on quantum phases, we compute the connected spin-spin correlation function and SPT string order parameter, as defined in Sec.~IV and ~V of the SM. It is important to note that LR interactions do not significantly affect the AFM long-range order for $\lambda < \lambda_{\rm c}$, as demonstrated in Sec.~V of the SM and previous works~\cite{Vodola_2016}. Therefore, in this work, we primarily focus on the influence of LR interactions on topological nontrivial QCPs and the gapped SPT phase.

In the SR limit $\alpha \rightarrow \infty$, our numerical results show AFM and cluster SPT phases for $\lambda < \lambda_{\rm c} = 0.5$ and $\lambda > \lambda_{\rm c}$, respectively, in agreement with previous studies~\cite{Son_2011,verresen2017prb,yu2024quantum} (see Sec.~III of the SM). In the presence of LR interactions, the simulation also verifies the stability of the gapped SPT phase. Fig.~\ref{fig2}~(a) and (b) depict a double logarithmic plot of the connected spin-spin correlation function $|C^{zz}_{\rm c}(r)|$ and string order parameter $\mathcal{S}(r)$ versus lattice distance $r$ for various $\alpha$ with fixed $\lambda=0.55$ within the gapped SPT region, revealing anomalous behaviors of the correlation function. Specifically, the connected spin-spin correlation displays a hybrid behavior that is exponential decay at short distances and algebraic decay at long distances for large $\alpha \textgreater 1$ and purely algebraic decay at all distances for sufficiently small $\alpha \lesssim 1$. On the other hand, the string order parameter displays ``long-range order'' and purely algebraic decay for large and small $\alpha$, respectively (see Sec.~V of the SM for more results). These properties are similar to critical states in SR interacting systems, but with a finite bulk energy gap $\Delta_{\rm b}$, as demonstrated in Sec.~IV of the SM for typical values of $\lambda \textgreater \lambda_{\rm c}$ where the cluster interaction dominates. 

To explicitly demonstrate the robustness of topological features for $\lambda > \lambda_{\rm c}$, we present the bulk entanglement spectrum (as defined in Sec.~III of the SM) and the magnetization profile $\langle\sigma^{z}_{i}\rangle$ for particular values of $\alpha$. For $\alpha \rightarrow \infty$, the ground state corresponds to the gapped SPT phase at $\lambda > \lambda_{\rm c}$, as evidenced by the four-fold degeneracy in the bulk entanglement spectrum and the obvious magnetization near the boundaries~\cite{pollmann2010prb,yu2024universal,scaffidi2017prx,yu2022prl,parker2018prb}. For finite $\alpha$, the four-fold degeneracy in the lowest-lying entanglement spectrum and the edge magnetization remain robust against LR interactions, as depicted in Fig.~\ref{fig2} (c) and (d) for $\alpha=0.3$ and $\lambda=0.55$ as an example. Consequently, within the sufficiently small $\alpha\lesssim 1$ regime, in addition to the existence of four-fold degenerate edge modes and a finite bulk energy gap, the ground state also displays algebraically decaying string and spin-spin correlations simultaneously. Therefore, the two key properties of the gapless topological phase, a degenerate entanglement spectrum and a power-law decay string order parameter (see Sec.~IV and ~V in the SM for details), are both observed. Hence, we term this phase as \emph{algebraic SPT} (the lower right corner of the phase diagram in Fig.~\ref{fig:phasediagram}), exhibiting nontrivial gapless topological behaviors even when the bulk is gapped. However, we want to emphasize that the algebraic behavior of the correlation functions solely arises from the LR interaction and does not relate to any gaplessness.

\emph{Topological Gaussian universality.}---After exploring the ground-state phase diagram, we now investigate the quantum phase transition between AFM and SPT phases for typical values of $\alpha$. To determine the transition point, we compute the Binder ratio for the AFM order~\cite{sandvik2010computational}, $R_{4} = \frac{1}{2}(3-\langle O^{4}_{\text{AFM}}\rangle / \langle O_{\text{AFM}}^{2} \rangle^{2})$ where $O_{\rm AFM} = \frac{2}{L} \sum\nolimits_{i=L/4+1}^{3L/4} (-1)^{i} \sigma_{i}^{z}$. In the long-range ordered phase, $R_{4}$ increases with the system size $L$. At the critical point, $R_{4}$ has a zero scaling dimension and is independent of $L$. In Fig.~\ref{fig3} (a) and (c), we present the results of $R_{4}$ versus $\lambda$ for $\alpha=3.0,0.6$, respectively (see Sec.~III of the SM for other $\alpha$); the crossing points pin down the AFM-SPT QCPs. We find the critical value $\lambda_{\rm c}$ of the transition decreases upon decreasing LR power exponent $\alpha$. Intuitively, as the LR interaction decays more slowly (smaller $\alpha$), the Ising part of the Hamiltonian becomes more frustrated and the cluster part will be more favored. 
We notice that our approach does not allow us to extract the critical point when $\alpha \lesssim 0.1$, since within our numerical results, Binder ratio with different $L$ do not cross at a single point in this region. Furthermore, we find that the conformal symmetry is preserved and broken for large and sufficiently small $\alpha$, respectively, as shown in Sec.~VI of the SM. 
Nevertheless, we can still define an effective central charge via $S(l) = \frac{c_{\rm eff}}{3} \ln[L/\pi \sin{(l\pi/L)}] + S_{0}$ where $S(l)$ is the entanglement entropy for an interval of length $l$, for the non-conformal region to characterize the universality class of the LR model~\cite{nishimoto2011prb} (see Sec.~II of the SM for details). 
Fig.~\ref{fig3} (b) and (d) exhibit $c_{\rm eff}$ as a function of $\lambda$ with fixed $\alpha$ displaying a sharp peak at the critical point, which is consistent with the results given by the Binder ratio in Fig.~\ref{fig3} (a) and (c). In the following, the topological features of the critical line will be studied at the critical points determined from $c_{\rm eff}$ since the finite-size effect is generally less severe under PBC compared with OBC.

\begin{figure}[tb]
    \includegraphics[width=0.9\linewidth]{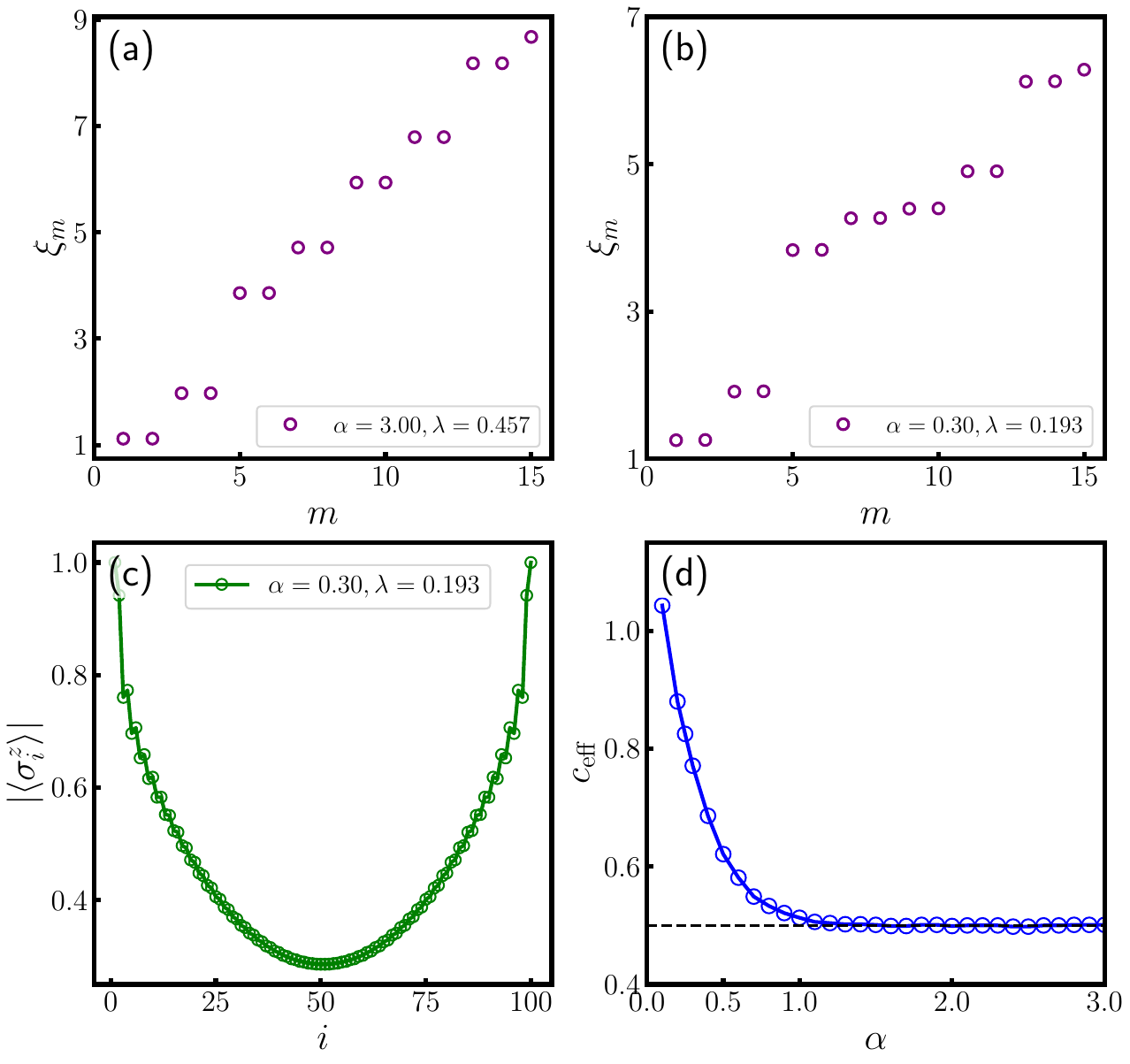}
    \caption{The bulk entanglement spectrum $\xi_{m}$ at the critical points for $\alpha=3.0$ (a) and $\alpha=0.3$ (b). The index $m$ counts the spectrum from the lowest-lying levels. (c) The magnetization profile at the critical point for a sufficiently small $\alpha=0.3$\,. (d) The effective central charge along the critical line as a function of $\alpha$ for a fixed system size $L=96$ under PBC.}
    \label{fig:fig4}
\end{figure}

More importantly, since the critical point of the SR cluster Ising model exhibits nontrivial topological properties such as degenerate edge modes, it is natural to examine whether these properties are robust against the LR interaction. Specifically, we demonstrate the two-fold topological degeneracy by computing the bulk entanglement spectrum at the critical point for $\alpha=3.0$ and $0.3$, respectively, as shown in Fig.~\ref{fig:fig4} (a) and (b). Similarly, Fig.~\ref{fig:fig4} (c) depicts that the critical edge mode is stable even for sufficiently small $\alpha = 0.3$ (see Sec.~III of the SM for other $\alpha$). This indicates that the entire critical line in the phase diagram exhibits nontrivial topological properties, which are also verified by the anomalous scaling of the edge mode energy splitting (see Sec.~VI of the SM for details). More remarkably, we present the results of the effective central charge $c_{\text{eff}}$ versus LR power exponent $\alpha$ in Fig.~\ref{fig:fig4} (d). The result unambiguously shows that within the numerical accuracy, the value of $c_{\text{eff}}$ tends to $1$ as $\alpha \approx 0.1$, corresponding to a Gaussian universality with nontrivial gapless topological behaviors. Hence, we term it as \emph{topological Gaussian universality}, as depicted by the yellow dashed line in the phase diagram in Fig.~\ref{fig:phasediagram}. Despite the numerical difficulties for $\alpha < 0.1$, fortunately, for $\alpha=0$, the 1D geometry of the system is completely lost and this limit case can be studied analytically (See Sec.~VII of the SM for details). Consequently, we find that the LR interaction induces a new gapped topological phase with a two-fold degenerate edge mode, which occupies the entire interaction region ($\lambda \textgreater 0$) for $\alpha=0$ (the red solid line in Fig.~\ref{fig:phasediagram}).

\emph{Discussion.}---To gain a physical understanding of the numerical results presented above, we first focus on the effect of the LR interaction on the bulk. The Hamiltonian \eqref{E1} can be transformed into the LR transverse field Ising model by the unitary $U=\prod_i CZ_{i,i+1}$. Under the renormalization group flow, the LR interaction will flow to operators in the infrared continuum field theory and correspond to both a local part $S_L$ and a non-local part $S_H$. Depending on $\alpha$, the non-local part can be irrelevant for large $\alpha$ or relevant for small $\alpha$~\cite{defenu2023rmp,yu2023prb} at the criticality. When the non-local part flows strong, the conformal symmetry will not emerge in the infrared. Away from the criticality, the competition between the local and non-local parts contributes to the hybrid behavior in the correlation function for any finite $\alpha$, as argued in~\cite{patrick2017prl,LEPORI201635}. Since the effective central charge is defined via the entanglement entropy, the non-local interaction will create more entanglement between the disjoint regions. Therefore, less $\alpha$ will increase $c_{\text{eff}}$~\cite{latorre2005pra}, eventually reaching $1.0$ as $\alpha$ approaches $0$~\cite{LEPORI201635}.

It is instructive to understand the case when $\alpha = 0$, where the LR interaction becomes all-to-all interaction. In this case, the system becomes effectively a ($0+1$)D system and has a large site permutation symmetry. As a result, all the correlation functions $\langle \sigma_i^z \sigma_j^z\rangle$ are constants depending only on $\lambda$. The gap size remains constant, corresponding to dynamical exponent $z=0$. Thus, the ground state represents a gapped topological phase characterized by two-fold degenerate edge modes, as illustrated in the next paragraph.

To verify the stability of critical edge modes against LR interactions, let's consider the open finite chain with size $L$. The local operators $\sigma_1^z,\sigma_L^z$ commute with the Hamiltonian, and the four-fold degenerate ground states are given by,
\begin{equation}
    \ket{\uparrow_L\uparrow_R} \pm \ket{\downarrow_L\downarrow_R}, \quad \ket{\uparrow_L\downarrow_R} \pm \ket{\downarrow_L\uparrow_R} \,,
\end{equation}
where $L(R)$ denotes the leftmost (rightmost) site. These four edge modes are robust in the SPT phase with LR interactions except for $\alpha = 0$. For $\alpha= 0$, the all-to-all interaction contains the term $\sigma_1^z \sigma_L^z$ which will favor the antiferromagnetic combination $\ket{\uparrow_L\downarrow_R} \pm \ket{\downarrow_L\uparrow_R}$. For general $\alpha$, when driving the system to the critical line, the Ising interaction no matter SR or LR will generate the $\sigma_1^z \sigma_L^z$ coupling, which results in the two-fold degenerate ground state, $\ket{\uparrow_L\downarrow_R} \pm \ket{\downarrow_L\uparrow_R}$. In short, the critical line with varying $\alpha$ exhibits robust two-fold degenerate edge modes, regardless of the LR interaction. The SPT phases with finite $\alpha$ have four-fold degenerate edge modes, but it is lifted to two-fold degenerate edge modes when the LR interaction is all-to-all (see Sec.~VII of the SM for details).

\emph{Concluding remarks.}---In summary, we have conducted large-scale numerical simulations to demonstrate the ground-state properties of a spin-1/2 chain with LR interactions, unveiling various exotic phases and phase transitions as a function of the cluster interaction strength $\lambda$ and the LR power exponent $\alpha$. For large $\alpha \textgreater 1.0$, the phase diagram exhibits the AFM long-range order, the gapped SPT phase, and a symmetry-enriched Ising criticality between them. Conversely, the results for sufficiently small but finite values of $\alpha \lesssim 1.0$ are particularly intriguing. The conneted spin-spin correlation function and string order parameter gradually change to purely algebraic decay for $\lambda \textgreater \lambda_{\rm c}$, implying the emergence of an algebraic SPT phase with nontrivial gapless topological behaviors. Furthermore, in the limit $\alpha = 0$, the entire interaction region is dominated by a distinct gapped topological phase with two-fold degenerate edge modes. More remarkably, the universality class gradually changes from topological Ising to the topological Gaussian universality as $\alpha$ approaches $0$, indicating that LR interactions can induce a new topological universality class. Generalization of these findings for higher dimensions is a fascinating and challenging direction for future study. From the perspective of experimental realizations, LR interacting spin chains hold the potential for implementation in state-of-the-art quantum circuit platforms~\cite{chen2023scipost,shen2023observation}. It has been proposed that highly entangled quantum matter and unconventional phase transitions can be observed in quantum circuits~\cite{iqbal2024non,iqbal2023topological,nat2023prl,zhu2023prl,verresen2022efficiently,tantivasadakarn2022longrange,zhu2023structured,sang2021prr,lavasani2022monitored,lavasani2021measurement,morral2023prb}, which may be useful to demonstrate our findings experimentally (details are discussed in Sec.~VIII of the SM). Hence, our findings open a new avenue toward exploring and understanding gapless topological phases of matter.

\textit{Acknowledgement}: We thank Da-Chuan Lu, Ruizhe Shen and Youjin Deng for helpful discussions. Numerical simulations were carried out with the ITENSOR package~\cite{itensor} on the Kirin No.2 High Performance Cluster supported by the Institute for Fusion Theory and Simulation (IFTS) at Zhejiang University. X.-J.Yu is supported by a start-up grant XRC-23102 of Fuzhou University. This work is also supported by MOST 2022YFA1402701.

\bibliographystyle{apsrev4-2}

\bibliography{main}

\clearpage
\onecolumngrid
\newpage

\def\normOrd#1{\mathop{:}\nolimits\!#1\!\mathop{:}\nolimits}
\renewcommand{\(}{\left(}
\renewcommand{\)}{\right)}
\renewcommand{\[}{\left[}
\renewcommand{\]}{\right]}

\def\Eq#1{Eq.~(\ref{#1})}
\def\Fig#1{Fig.~\ref{#1}}
\def\avg#1{\left\langle#1\right\rangle}

\section{Supplemental Material for Gifts from long-range interaction: Emergent gapless topological behaviors in quantum spin chain}

\subsection{Section I: Density matrix renormalization group algorithm and numerical setups}
\label{sm:dmrg}

To determine the ground state of the long-range (LR) interacting cluster Ising model introduced in the main text, we conducted extensive density matrix renormalization group (DMRG) simulations~\cite{white1992prl,white1993prb,SCHOLLWOCK201196,schollwock2005rmp} using the matrix product state (MPS) representation~\cite{Cirac2006PRB}. 
Over the past few decades, DMRG has emerged as a powerful tool for studying one-dimensional (1D) quantum many-body systems with local interactions. 
However, when dealing with 1D quantum systems featuring LR interactions, the ground state no longer necessarily adheres to the so-called entanglement area law~\cite{Eisert2010RMP}, which can hinder an efficient MPS representation. 
In our specific scenario, the ground-state phase diagram (refer to Fig.~1 in the main text) exhibits area-law or logarithmic entangled phases (as well as critical points). 
Despite the complexity induced by LR interactions, a faithful MPS representation remains feasible by employing a sufficiently large MPS bond dimension for finite-size systems.

\begin{figure}[tb]
    \includegraphics[width=0.9\linewidth]{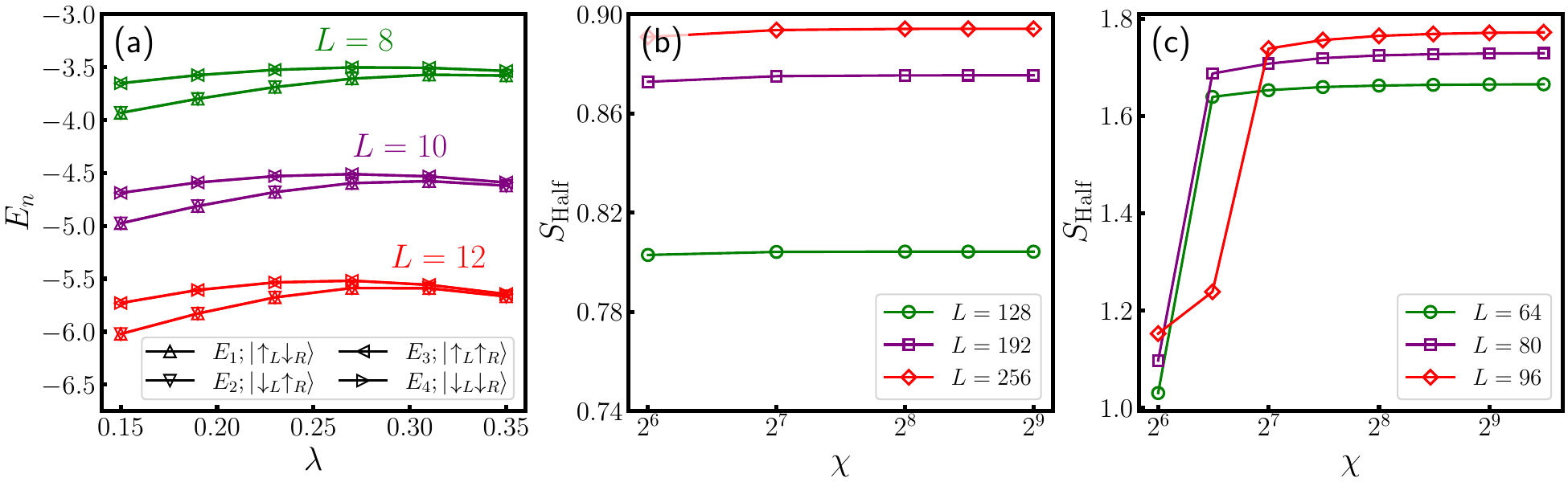}
    \caption{(a) ED solution of the first four lowest-lying energy levels of the LR cluster Ising model for $L=8$, $10$, and $12$ along the $\alpha=0.5$ line under open boundary conditions. The ground state is doubly degenerate with the two boundary spins pointing to anti-parallel directions, namely, $\ket{\uparrow_{L}\downarrow_{R}}$ and $\ket{\downarrow_{L}\uparrow_{R}}$. The convergence of the half-chain entanglement entropy $S_{\rm Half}$ with respect to the maximum MPS bond dimension $\chi$ for typical system sizes under (b) open and (c) periodic boundary conditions, respectively. An MPS bond dimension $\chi = 512$ is sufficiently large for the results to be converged in DMRG calculations. Here, the simulated parameter is $\alpha=0.5$ and $\lambda=0.253$\,.}
    \label{fig:benchmark}
\end{figure}

In practice, we simulated finite systems of an even total length, i.e., $L \in 2\mathbb{Z}^{+}$, under open or periodic boundary conditions with a maximum MPS bond dimension set to $512$. 
Once the MPS energy has converged up to the order $10^{-8}$, the DMRG sweeping route would be stopped; 
the resulting MPS can be treated as a faithful representation of the true ground state. 
It is also noted that the initial MPS is chosen as one of the N\'{e}el states for two reasons: 
i) for open boundary conditions, there is no spin-flip operator $\sigma^{x}$ in the Hamiltonian acting on the leftmost ($L$) or rightmost ($R$) boundary site, it is thus necessary to set the two boundary spins to anti-parallel directions, namely, $\ket{\uparrow_{L} \downarrow_{R}}$ or $\ket{\downarrow_{L} \uparrow_{R}}$, so that the calculation is converged to the correct result. 
In Fig.~\ref{fig:benchmark} (a), we have verified that the doubly degenerate ground states indeed have two anti-parallel directed boundary spins via the calculation of the exact diagonalization (ED).
Of course, one can also choose the product state $\prod_{i=1}^{L} \ket{+}_{i}$, in which $\sigma_{i}^{x} \ket{+}_{i} = \ket{+}_{i}$, as the starting point. 
The product state respects the global symmetry $G = \prod_{i=1}^{L} \sigma_{i}^{x}$ and finally converges to a superposition $\ket{\uparrow_{L} \downarrow_{R}} + \ket{\downarrow_{L} \uparrow_{R}}$. 
Compared to the former choice, the latter one usually requires a larger MPS bond dimension to achieve the same accuracy level. 
ii) for periodic boundary conditions, the main quantity under investigation is the entanglement entropy and its spectrum. 
More specifically, the nontrivial topological property of the critical point is partly identified by the double degeneracy of the lowest-lying level of the entanglement spectrum. 
If the MPS is initialized to be the product state $\prod_{i=1}^{L} \ket{+}_{i}$, a doubly degenerate entanglement spectrum would be obtained within the anti-ferromagnetic (AFM) phase (spontaneously symmetry breaking happens only in the thermodynamic limit). 
To have a clear distinction between the AFM phase, cluster symmetry-protected-topological (SPT) phase, and the nontrivial critical point, it is useful to artificially break the $G$ symmetry at the beginning. 
By initializing the MPS as one of the N\'{e}el states, the resulting degeneracy of the lowest-lying level of the entanglement spectrum would be 1, 2, and 4 for the AFM phase, topological critical point, and cluster SPT phase, respectively.

\begin{figure}[tb]
    \includegraphics[width=0.65\linewidth]{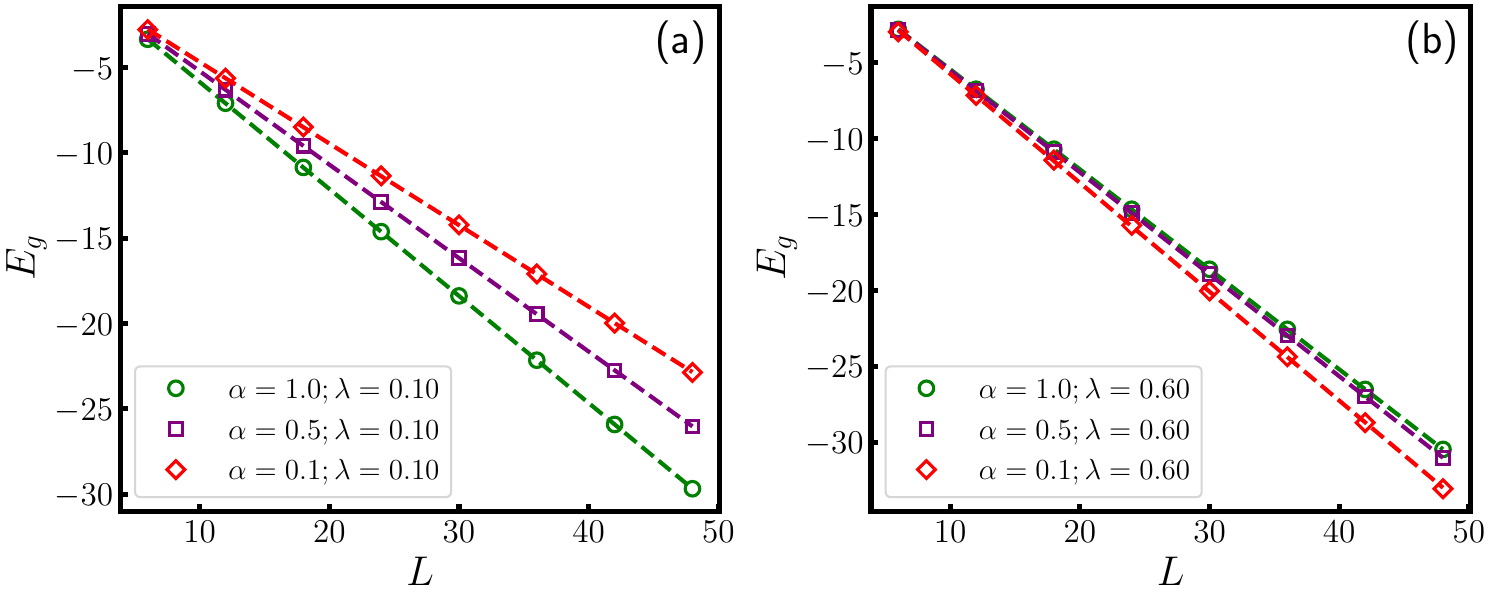}
    \caption{The ground-state energy $E_{g}$ of the LR cluster Ising model under the open boundary condition as a function of the system size for typical values of $\alpha$ with (a) $\lambda=0.1$ and (b) $\lambda=0.6$, respectively. The open markers are numerical data and the dashed lines are linear least-squares fittings.}
    \label{fig:nokacfactor}
\end{figure}

Additionally, we would like to note that it is common to include a Kac factor as a coefficient of the LR ferromagnetic interaction to preserve the Hamiltonian extensive. 
In the present scenario, however, the LR interaction is anti-ferromagnetic (or frustrated). 
It means it is impossible to simultaneously minimize all LR terms $\sigma_{i}^{z} \sigma_{j}^{z} / d_{ij}$, therefore, the ground-state energy of the LR cluster Ising Hamiltonian can still be an extensive quantity which is linear with the size $L$. 
As illustrated in Fig.~\ref{fig:nokacfactor}, we have numerically verified a clear linear dependence of the ground-state energy $E_{g}$ with respect to the system size $L$ for sufficiently small $\alpha \lesssim 1.0$\,. 
Based on this observation, the Kac factor is not necessary in our case.

Finally, we want to demonstrate that an MPS bond dimension $\chi = 512$ is sufficiently large in our studies.
Since the half-chain entanglement entropy $S_{\rm Half}$ is usually the quantity slowest to converge in DMRG calculations, we have also tested the convergence of $S_{\rm Half}$ with respect to the bond dimension $\chi$ as illustrated in Fig.~\ref{fig:benchmark} (b) and (c).
It is confirmed that $\chi=512$ can give converged reliable results.

To explore the interesting physics brought by the LR interactions, we have calculated various quantities, such as the connected spin-spin correlation, the energy gap, the entanglement spectrum, and so on, which we detail in the following sections.

\subsection{Section II: Logarithmic entanglement entropy and effective central charge}
\label{sm:charge}

It is now well accepted that quantum entanglement plays an essential role in our understanding of quantum many-body physics. 
In particular, the low-energy eigenstates of a gapped Hamiltonian with local interactions are known to obey the entanglement area law~\cite{Eisert2010RMP}, indicating that the entanglement entropy for an interval of a 1D chain is constant regardless of the subsystem size $l$, namely, $S(l) \equiv - \text{tr} [\rho_{l} \log(\rho_{l})] \sim \text{const}$, where $\rho_{l}$ is the reduced density matrix of the subsystem. 
The entanglement area law explains why MPS works so well for solving 1D quantum many-body problems. 

On the other hand, the entanglement entropy also provides a convenient way to extract universal characteristics of quantum criticality. 
For a (1+1)D critical system described by a conformal field theory (CFT), the central charge $c$ of the underlying CFT can be determined from the universal logarithmic scaling of the entanglement entropy via the famous formula~\cite{Holzhey1994NPB,Calabrese2004JSM,Calabrese2009JPA}
\begin{equation}
    \label{eq:charge}
    S(l) = \frac{c}{3} \log{ \left[ \frac{L}{\pi} \sin\left( \frac{\pi l}{L} \right) \right] } + S_{0} \, ,
\end{equation}
where $l$ the interval length, $L$ the system total size, and $S_{0}$ a non-universal constant; 
$L/\pi \sin(\pi l/L)$ is also called the chord length. 
The distinct scaling behaviors of $S(l)$ shown in the gapped phase and at the critical point give us an effective way to pin down the location of quantum critical points.

\begin{figure}[tb]
    \includegraphics[width=0.85\linewidth]{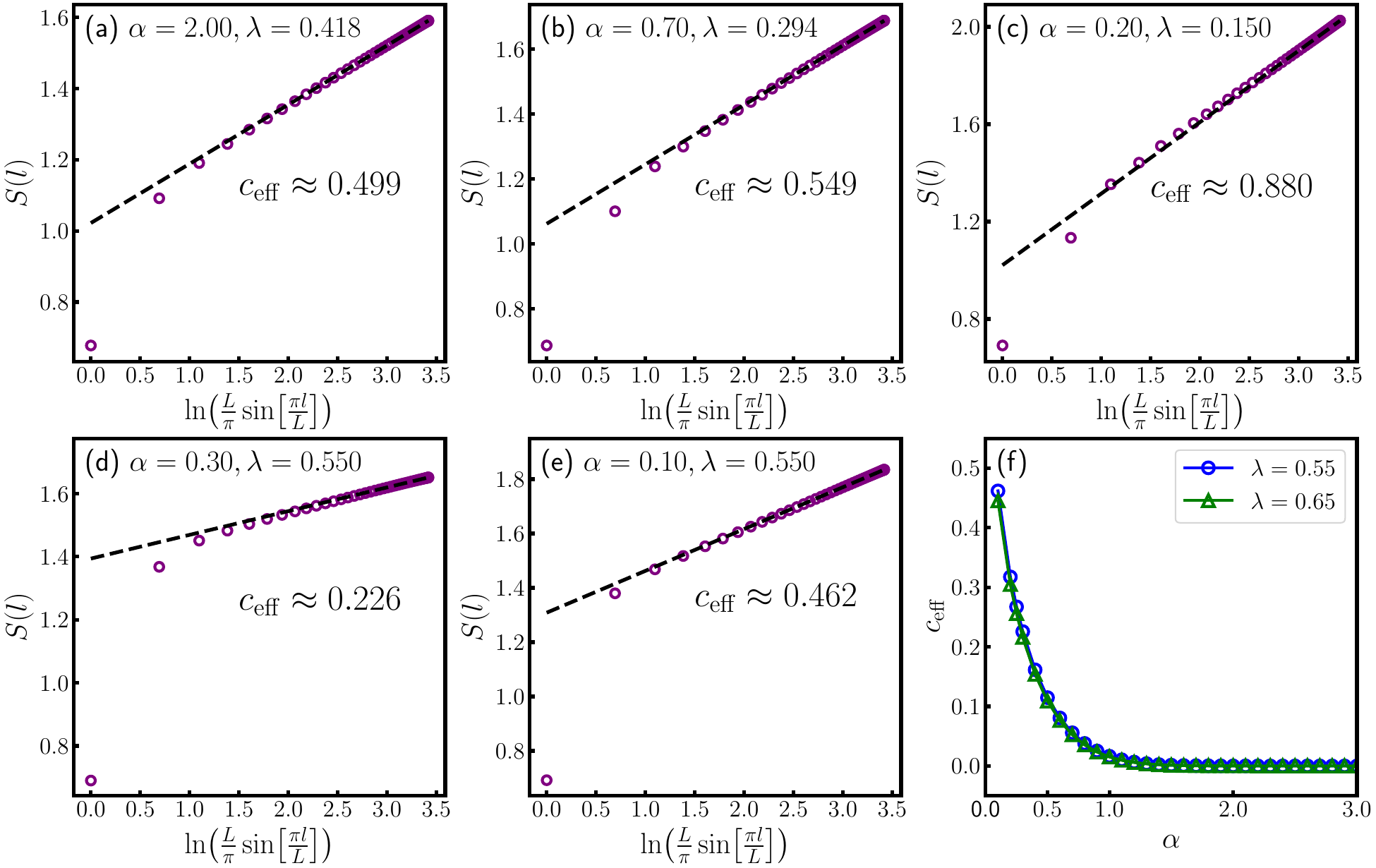}
    \caption{The entanglement entropy shows a logarithmic scaling with the chord length (a)-(c) along the critical line and (d)-(e) within the gapped SPT phase for sufficiently small $\alpha \lesssim 1$. (f) summarizes the effective central charge as a function of $\alpha$ along $\lambda=0.55$ and $0.65$ lines. Purple circles represent DMRG data and the dashed lines are least-squares fittings according to Eq.~\eqref{eq:charge}. The simulated system size is $L=96$ under the periodic boundary condition for all $\alpha$ and $\lambda$.}
    \label{fig:charge}
\end{figure}

In the present work, we simulated systems of size $L=96$ under periodic boundary conditions to calculate $S(l)$ where $l \leq L/2$; 
a least-squares fitting according to Eq.~\eqref{eq:charge} is then performed to extract the coefficient of the logarithmic scaling. 
By scanning through the whole ground-state phase diagram, in addition to the critical points, we also found the entanglement entropy display a logarithmic scaling with the chord length within the gapped SPT phase for sufficiently small $\alpha \lesssim 1$; 
see Fig.~\ref{fig:charge} (d) and (e) for two representative examples. 
This implies that the LR interaction can break down the entanglement area law obeyed by gapped phases. 
However, the violation of entanglement area law is only observed within the SPT phase; 
the AFM phase still satisfies the area law. 
Since the breakdown of the area law is found to be logarithmic, we can thus define an effective central charge $c_{\rm eff}$ to characterize the entanglement scaling behaviors within the whole phase diagram. 
It is also worth to emphasize that the value of $c_{\rm eff}$ extracted from the formula~\eqref{eq:charge} does not necessarily relate to a CFT. 
In our case, $c_{\rm eff}$ corresponds to the true central charge of an Ising CFT only along the critical line with $\alpha \gtrsim 1$; for $\alpha \lesssim 1$, the conformal symmetry breaks down with a dynamical exponent $z<1$ and the central charge is not well-defined in the low-energy field theories describing the critical points. 
At last, Fig.~\ref{fig:charge} gives some representative results to support the observation made above; 
the critical point $\lambda_{\rm c}$ for a given $\alpha$ used in Fig.~\ref{fig:charge} (a), (b), and (c) is identified by the peak position of $c_{\rm eff}$ as a function of $\lambda$ along the fixed $\alpha$ line [see Fig.~\ref{fig3} (b) and (d) in the main text for examples].

\subsection{Section III: Numerical results of critical and topological properties at quantum critical points for other LR power exponents $\alpha$}
\label{sm:critical_line}

In the main text, we presented extensive numerical results for $\alpha=3.0$ and $0.6$, which reveal nontrivial topological features at quantum critical points between the AFM and cluster SPT phases. 
In this section, to further corroborate our conclusions and investigate how the topological stability of the critical line evolves when decreasing $\alpha$, we performed simulations to study the model with other LR power exponents $\alpha$. 
As mentioned above, the critical point $\lambda_{c}$ can be determined by the peak position of the effective central charge $c_{\rm eff}$ as a function of $\lambda$ along the fixed $\alpha$ line. 
Here, we employed another commonly used quantity, the Binder ratio~\cite{binder1981}, to provide evidence that there remains to be a phase transition between the AFM and cluster SPT phases as $\alpha$ decreases. 
Moreover, the critical point obtained independently from the Binder ratio can serve as a cross-check for the correctness of $\lambda_{c}$ determined from the effective central charge.

The Binder ratio is a useful tool in the study of different types of quantum phase transitions. 
With the definition of the AFM order parameter
\begin{equation}
    O_{\rm AFM} = \frac{2}{L} \sum\nolimits_{i=L/4+1}\nolimits^{3L/4} (-1)^{i} \sigma_{i}^{z}
\end{equation}
where we have restricted the sum over the middle region of the chain to reduce boundary effects, the corresponding Binder ratio is given by 
\begin{equation}
    R_{4} = \frac{1}{2} \left( 3 - \frac{\langle O_{\rm AFM}^{4} \rangle}{\langle O_{\rm AFM}^{2} \rangle^{2}} \right) \, .
\end{equation}
This observable has a vanishing scaling dimension and can therefore provide an unbiased estimation of the critical point. 
In addition, we calculated the long-range order relevant to the cluster SPT phase, which equals the product of cluster terms~\cite{verresen2021prx}, 
\begin{equation}
    O_{\rm SPT} = \sum\nolimits_{i=L/4+1}\nolimits^{3L/4} \sigma_{i-1}^{z} \sigma_{i}^{x} \sigma_{i+1}^{z} = \sigma_{L/4}^{z} \sigma_{L/4+1}^{y} \left(\prod\nolimits_{j=L/4+2}\nolimits^{3L/4-1} \sigma_{j}^{x} \right) \sigma_{3L/4}^{y} \sigma_{3L/4+1}^{z} \, .
\end{equation}

\begin{figure}[tb]
    \includegraphics[width=1.0\linewidth]{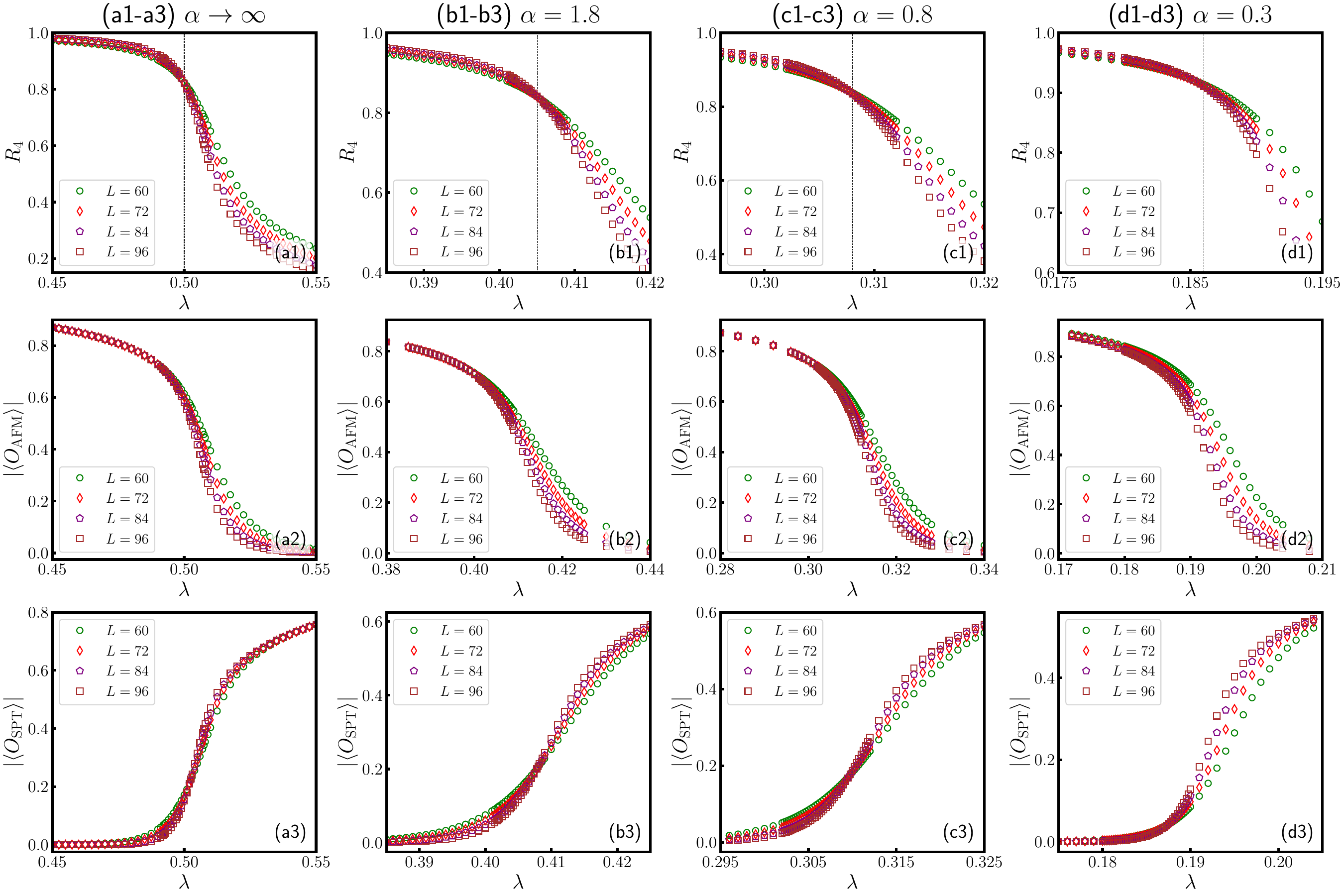}
    \caption{The Binder ratio of the AFM order $R_{4}$ (the upper panel), the AFM order parameter $\langle{O_{\rm AFM}}\rangle$ (the middle panel), and the SPT string order parameter $\langle{O_{\rm SPT}}\rangle$ (the lower panel) as a function of $\lambda$ are shown for (a1-a3) $\alpha \rightarrow \infty$, (b1-b3) $\alpha=1.8$, (c1-c3) $\alpha=0.8$, and (d1-d3) $\alpha=0.3$, with system sizes $L=60$, $72$, $84$, and $96$ under open boundary conditions. The critical points determined through the Binder ratio crossings are marked by black dashed vertical lines in the upper panel.}
    \label{fig:binder_order}
\end{figure}

As displayed in the upper panel of Fig.~\ref{fig:binder_order}, the Binder ratio of different system size $L$ intersect with each other at a single point giving an estimation of the critical point $\lambda_{c}$ consistent with the one obtained from the effective central charge (see Fig.~1 in the main text for details).
Similarly, the results of $|\langle{O_{\rm AFM}}\rangle|$ and $|\langle{O_{\rm SPT}}\rangle|$ explicitly show that the ground states are AFM and cluster SPT phases for $\lambda < \lambda_{c}$ and $\lambda > \lambda_{c}$, respectively. 
The quantum phase transition between the two phases is ubiquitously present at all values of $\alpha$; the critical point shifts to a smaller value as expected due to the enhancement of anti-ferromagnetic frustrations as $\alpha$ decreases.

\begin{figure}[tb]
    \includegraphics[width=0.85\linewidth]{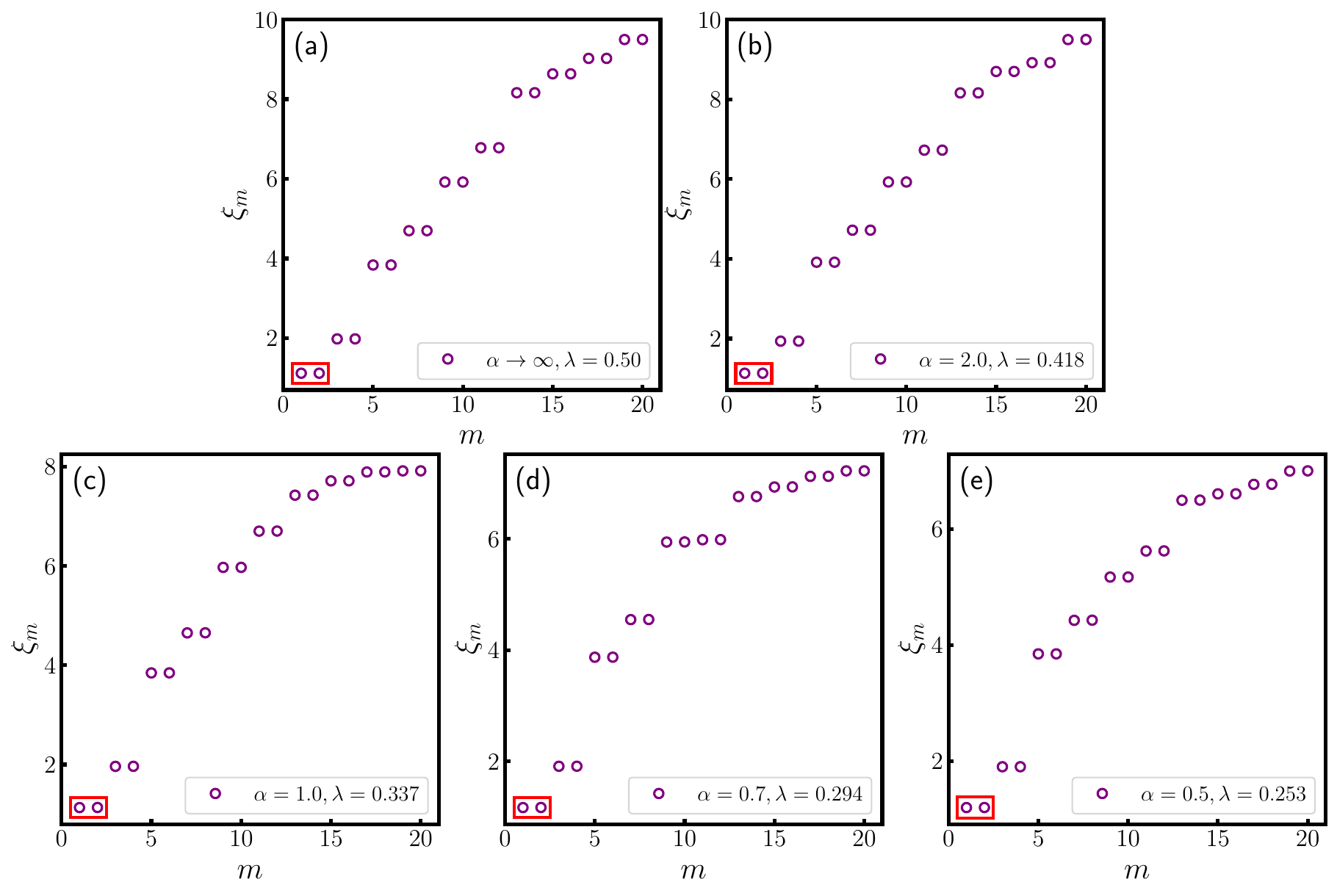}
    \caption{The bulk entanglement spectrum $\{\xi_{m}\}$ for fixed system size $L=96$ under periodic boundary conditions at the critical line is displayed for (a) $\alpha \rightarrow \infty$, (b) $\alpha=2.0$, (c) $\alpha=1.0$, (d) $\alpha=0.7$, and (e) $\alpha=0.5$\,. The index $m$ counts the spectrum from the lowest-lying levels and only the first $20$ levels are presented in the plot. The red boxes highlight the two-fold degeneracy in the entanglement spectrum.}
    \label{fig:spectrum_cp}
\end{figure}

\begin{figure}[tb]
    \includegraphics[width=0.7\linewidth]{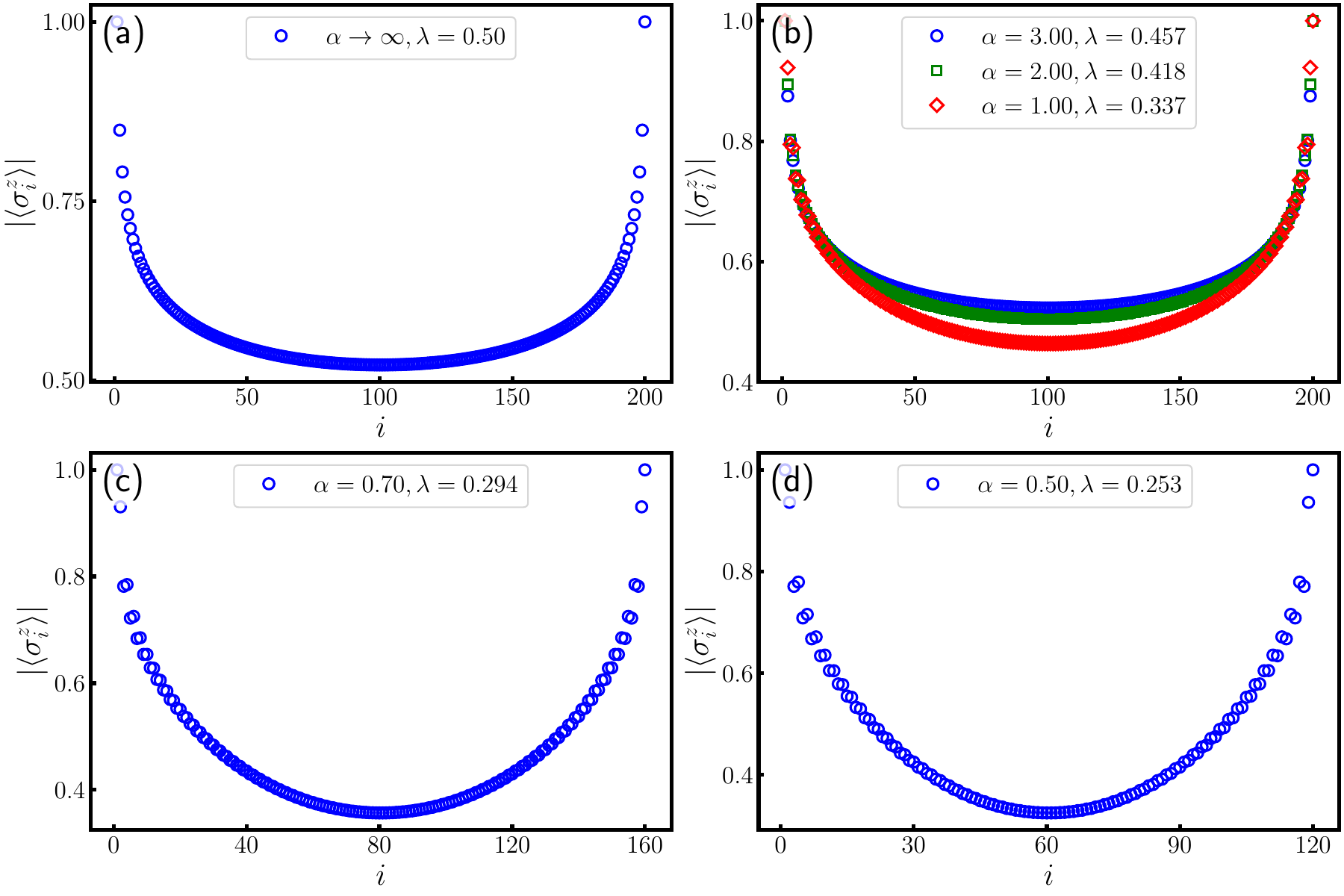}
    \caption{The magnetization profile $\langle{\sigma_{i}^{z}}\rangle$ under open boundary conditions at the critical points (determined by the effective central charge $c_{\rm eff}$) is shown for (a) $\alpha \rightarrow \infty$, (b) $\alpha=3.0, 2.0, 1.0$, (c) $\alpha=0.7$, and (d) $\alpha=0.5$\,. 
    }
    \label{fig:profile_cp}
\end{figure}

To further verify the topological features of the critical line, we also investigate the bulk entanglement spectrum (under periodic boundary conditions) and magnetization profile (under open boundary conditions) at the critical point for various $\alpha$. 
The critical points explored here are determined by the effective central charge $c_{\rm eff}$. 
By partitioning the whole spin chain into two equal parts, i.e., $A = (1, \dots L/2)$ and $B = (L/2+1, \dots L)$, the so-called entanglement Hamiltonian $\tilde{H}_{A}$ is related to the half-chain reduced density matrix $\rho_{A} = \text{tr}_{B}(\ket{\psi}\!\bra{\psi})$ via $\tilde{H}_{\rm A} \equiv - \ln{\rho_{A}}$, where $\ket{\psi}$ is the ground-state wavefunction. 
The entanglement spectrum $\{\xi_{m}\}$ is then the eigenvalues of the entanglement Hamiltonian sorted in ascending order. 
According to the universal bulk-boundary correspondence, the degeneracy of the lowest-lying level in the entanglement spectrum can faithfully reflect the existence of edge modes living on the boundary. 
As evidenced by Figs.~\ref{fig:spectrum_cp} and~\ref{fig:profile_cp}, the profile $\langle{\sigma_{i}^{z}}\rangle$ shows an obvious magnetization near the open boundary and the lowest-lying level of the entanglement spectrum exhibits a clear double degeneracy, both indicating that the topological properties of the critical line remain stable against the LR AFM interactions.

\subsection{Section IV: Finite bulk energy gap in the algebraic SPT phase}
\label{sm:spt_gap}

In the main text, we have presented the results of the connected spin-spin correlation function and the nonlocal SPT string order parameter, which exhibit purely algebraic decay behaviors at all distances for sufficiently small $\alpha \lesssim 1.0$ deep in the cluster SPT phase; 
it corresponds to the lower right part of the phase diagram in the main text, which is termed algebraic SPT phase there. 
Here, we would like to clarify that this algebraic decay behavior is caused by the LR AFM interaction and is not related to a critical phase. 
To show that the algebraic SPT phase is gapped, it is straightforward to check the existence of a finite bulk energy gap in the thermodynamic limit, which we detail below.

Before we show the numerical results, it is helpful to understand the lowest-lying energy spectrum in the short-range cluster SPT phase ($\alpha \to \infty$ and $1/2 < \lambda \leq 1$). 
If we consider the limit $\lambda = 1$, the ground-state manifold under the open boundary condition would be four-fold degenerate coming from the edge modes on the boundary. 
Since the two $\sigma_{1}^{z}$ and $\sigma_{L}^{z}$ on the boundary commutes with the Hamiltonian, we can label these four states by $\ket{\uparrow_{L} \uparrow_{R}}$, $\ket{\uparrow_{L} \downarrow_{R}}$, $\ket{\downarrow_{L} \uparrow_{R}}$, and $\ket{\downarrow_{L} \downarrow_{R}}$, where $L$ ($R$) denotes the left (right) boundary. 
By including a finite Ising interaction but still within the gapped SPT phase ($1/2 < \lambda < 1$), the AFM interaction will penalize the FM states splitting the original four-fold degenerate manifold into two doubly degenerate manifolds, i.e., $\{ \ket{\uparrow_{L} \downarrow_{R}}, \ket{\downarrow_{L} \uparrow_{R}} \}$ and $\{ \ket{\uparrow_{L} \uparrow_{R}}, \ket{\downarrow_{L} \downarrow_{R}} \}$; 
the former one has a slightly lower energy compared with the latter one. 
It is noted that the small gap opened by the Ising interaction is finite only for finite-size systems, the four-fold ground-state degeneracy is exact (restored) in the thermodynamic limit.

\begin{figure}[tb]
    \includegraphics[width=0.65\linewidth]{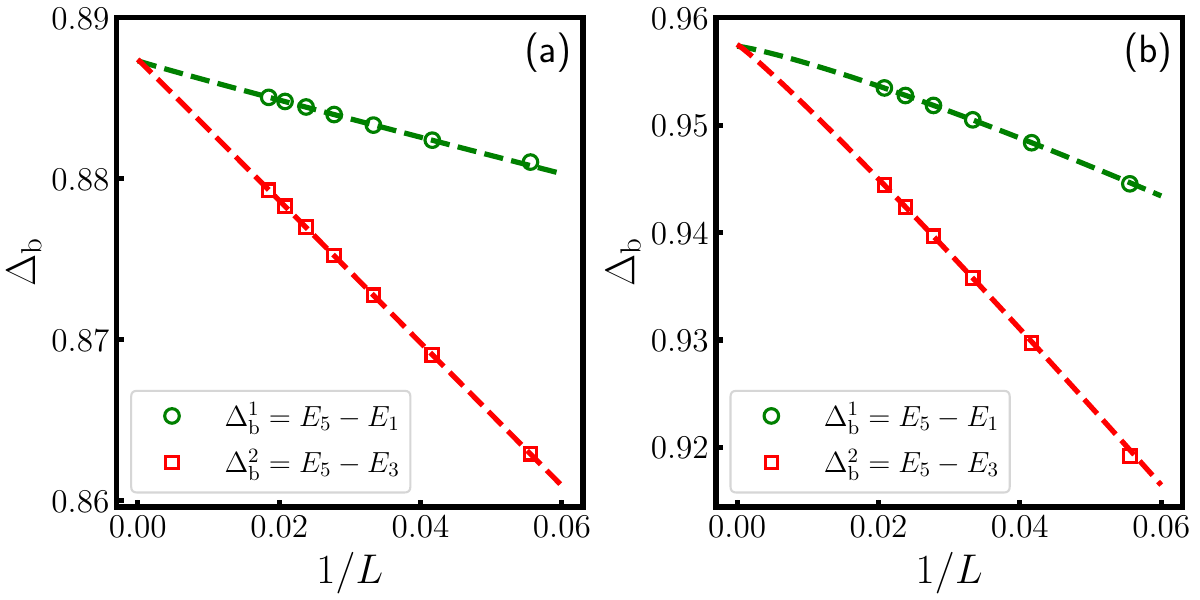}
    \caption{The bulk energy gap defined by $\Delta_{\rm b}^{1} \equiv E_{5} - E_{1}$ or $\Delta_{\rm b}^{2} \equiv E_{5} - E_{3}$ as a function of $1/L$ for (a) $\alpha=0.5, \lambda=0.60$ and (b) $\alpha=0.3, \lambda=0.60$\,. The dashed lines are least-squares fittings according to $\Delta_{\rm b}(L) = a / L^{b} + c$.}
    \label{fig:bulkgap_spt}
\end{figure}

Now we can consider the case of finite $\alpha$ and verify the finite bulk energy gap within the cluster SPT phase even for very small $\alpha \lesssim 1.0$\,. 
To this end, we have to calculate the first $6$ lowest-lying energy levels of the system to obtain the correct bulk energy gap; 
in this case, we initialized the MPS randomly in the DMRG simulations. 
By sorting the obtained energy levels in ascending order, we found a doubly degenerate energy spectrum, $E_{1,2} < E_{3,4} < E_{5,6}$, as expected. 
More specifically, the energy difference between $E_{1}$ and $E_{3}$ is much smaller than the one between $E_{3}$ and $E_{5}$. 
Based on this observation, the bulk energy gap can be defined separately by $\Delta_{\rm b}^{1} \equiv E_{5} - E_{1}$ and $\Delta_{\rm b}^{2} \equiv E_{5} - E_{3}$. 
As illustrated in Fig.~\ref{fig:bulkgap_spt}, $\Delta_{\rm b}^{1}$ and $\Delta_{\rm b}^{2}$ are extrapolated to the same value in the thermodynamic limit. 
It means that the first $4$ lowest-lying energy levels are degenerate in the $L \to \infty$ limit and there does exist a finite bulk energy gap in the algebraic SPT phase even when the LR interaction is very strong.

\begin{figure}[tb]
    \includegraphics[width=0.65\linewidth]{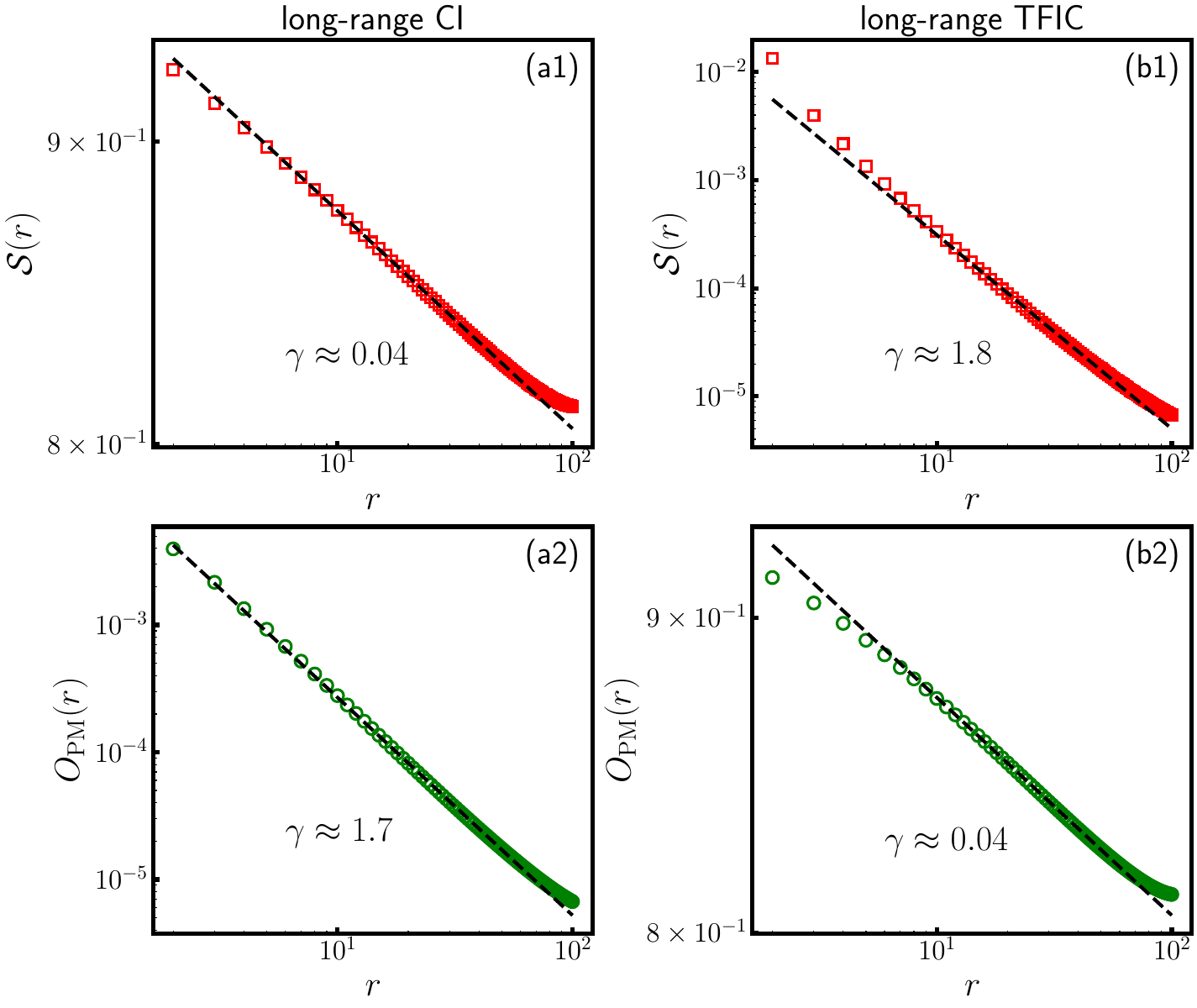}
    \caption{The log-log plot of the SPT string order parameter $\mathcal{S}(r)$ and the PM string order parameter $O_{\rm PM}(r)$ versus lattice distance $r$ at $\alpha = 0.3$ and $\lambda=0.55$, respectively, for LR cluster Ising model (the left panel) and LR transverse field Ising chain (the right panel). Simulated system size is $L=200$ and only middle sites are considered to reduce boundary effects. The dashed lines are least-squares fittings according to $O(r) = a/r^{\gamma}$, where $O(r)$ is $\mathcal{S}(r)$ or $O_{\rm PM}(r)$.}
    \label{fig:compare}
\end{figure}

In addition, we would like to mention that a similar phenomena has been observed in the LR transverse field Ising model, where there is a gapped paramagnetic (PM) phase featuring algebraic decaying correlations~\cite{Vodola_2016}. 
Although the algebraic SPT phase found here shares some similarities with the gapped PM2 phase reported in Ref.~\cite{Vodola_2016}, it is worth emphasizing that the algebraic SPT phase exhibits nontrivial gapless topological behaviors, such as the topological degeneracy in the bulk entanglement spectrum and localized magnetization near the boundaries. 
To have a direct comparison, we also computed two distinct types of nonlocal string orders
\begin{equation}
    \begin{split}
    & \mathcal{S}(r = |m-n|) = \langle \sigma^{z}_{n-1} \sigma^{y}_{n} \left( \prod\nolimits_{j=n+1}\nolimits^{m-1} \sigma^{x}_{j} \right) \sigma^{y}_{m} \sigma^{z}_{m+1}\rangle \, , \\
    & O_{\text{PM}}(r = |m-n|) = \langle \prod\nolimits_{j=n}\nolimits^{m} \sigma^{x}_{j}\rangle \, .
    \end{split}
    \label{eq:twostring}
\end{equation}
The former can be seen as the order parameter for the SPT phase, while the latter is the order parameter for the topologically trivial PM phase. 
In the short-range limit, i.e., $\alpha \to \infty$, $\mathcal{S}(r)$ ($O_{\rm PM}(r)$) is a long-range order within the SPT (PM) phase, while it decays exponentially fast in the PM (SPT) phase. 
By including a strong enough LR interaction ($\alpha \lesssim 1.0$), it is found that both $\mathcal{S}(r)$ and $O_{\rm PM}(r)$ become a quasi-long-range order exhibiting a power-law dependence on the string length, $\sim 1/r^{\gamma}$, whether the ground state is in the SPT or PM phase. 
However, the nontrivial gapless topological behaviors can be revealed through the slowest decay of the string operator, known as the symmetry-flux operator in the context of Ref.\cite{verresen2021prx}. In our case, the $\mathcal{S}(r)$ is charged under time-reversal symmetry, while the $O_{\text{PM}}(r)$ is not. Therefore, as evidenced by Fig.~\ref{fig:compare}, $\mathcal{S}(r)$ has a significantly smaller power exponent compared with $O_{\rm PM}(r)$ in the algebraic SPT phase. 
While in the PM2 phase, the situation is different, and $O_{\rm PM}(r)$ decays much slower than $\mathcal{S}(r)$. 
We note that the chosen parameters $\alpha=0.3$ and $\lambda=0.55$ correspond to the algebraic SPT phase in the LR cluster Ising chain and to the PM2 phase in the LR transverse field Ising chain. 
The direct comparison between the two models unequivocally reveals the topological distinction between the algebraic SPT phase found here and the PM2 phase reported in Ref.~\cite{Vodola_2016}.



\subsection{Section V: Asymptotic behavior of correlation functions}
\label{sec:SM5}

\begin{figure}[tb]
    \includegraphics[width=0.7\linewidth]{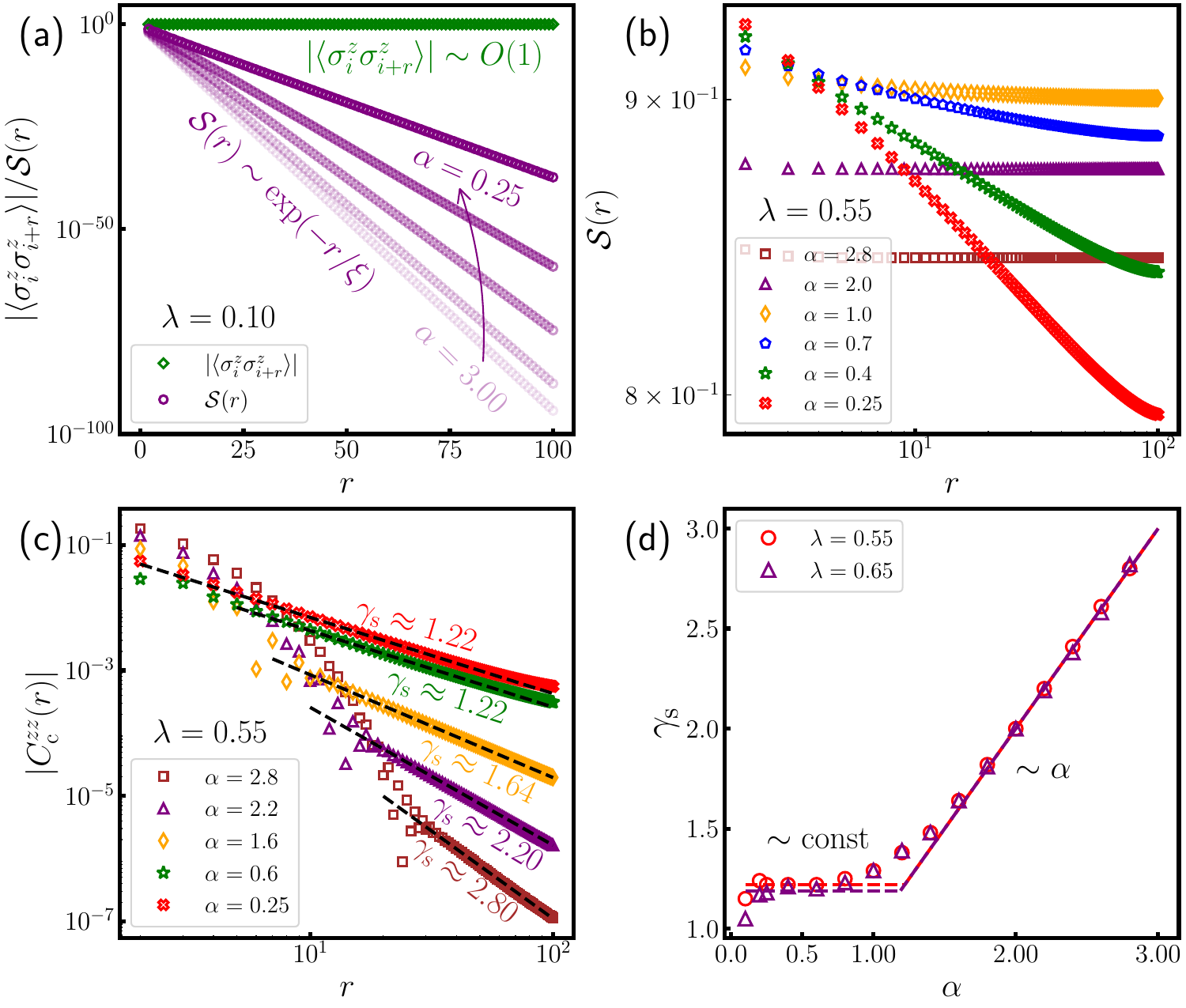}
    \caption{(a) In the AFM phase, $\vert{\langle{\sigma_{i}^{z}\sigma_{i+r}^{z}}}\rangle\vert \sim O(1)$ indicates the long-range AFM order, and $\mathcal{S}(r) \sim \exp(-r/\xi)$ decays exponentially. (b) In the SPT phase, the string order parameter $\mathcal{S}(r)$ changes from a long-range order $\sim O(1)$ to a power-law decay behavior as $\alpha$ decreases below $1.0$ with fixed $\lambda=0.55$\,. (c) The connected spin-spin correlation $|C_{\rm c}^{zz}(r)|$ for $\lambda=0.55$, showing a hybrid exponential and power-law decaying behavior when $\alpha \gtrsim 1.0$ and a purely algebraic decaying behavior when $\alpha \lesssim 1.0$\,. The black dashed lines are least-squares fittings according to $|C_{\rm c}^{zz}(r)| \sim 1/r^{\gamma_{\rm s}}$. (d) The power exponent $\gamma_{\rm s}$ as a function of $\alpha$, respectively, along the $\lambda=0.55$ (red circles) and $\lambda=0.65$ (purple triangles) lines. Simulated system size is $L=200$ under the open boundary condition. The finite-size effect is severe when $\alpha \lesssim 0.1$\,.}
    \label{fig:correlation}
\end{figure}

In this section, we examine various types of correlation functions in the presence of LR interactions. 
Specifically, we consider the connected spin-spin correlation function
\begin{equation}
    C_{\rm c}^{zz}(r) = \langle \sigma^{z}_{i}\sigma^{z}_{i+r} \rangle - \langle \sigma^{z}_{i} \rangle \langle \sigma^{z}_{i+r}\rangle
    \label{eq:czz}
\end{equation}
and the nonlocal SPT string order parameter $\mathcal{S}(r)$ [refer to Eq.~\eqref{eq:twostring} for its definition]. 
In the practical simulations, we have chosen $i=L/4$, and $r \in [1, L/2]$ to reduce the boundary effect.

As elaborated in the main text and above sections, the LR cluster Ising model exhibits a continuous phase transition between the gapped AFM and SPT phases by tuning the cluster interaction strength $\lambda$, irrespective of the magnitude of the LR interaction. 
In the AFM regime (for $\lambda$ below the critical point), we computed the spin-spin correlation function $\langle \sigma_{i}^{z} \sigma_{i+r}^{z} \rangle$ with respect to the lattice distance $r$ from large to small $\alpha$, as depicted in Fig.~\ref{fig:correlation}~(a). 
The result shows that the spin-spin correlation function exhibits an AFM long-range order, i.e., $|\langle \sigma_{i}^{z} \sigma_{i+r}^{z} \rangle| \sim O(1)$, for any $\alpha \textgreater 0$. 
Conversely, the nonlocal SPT order parameter always exhibits an exponential decay, $\mathcal{S}(r) \sim \text{exp}(-r/\xi)$, in which $\xi$ can be seen as the correlation length. 
For a given $\lambda$, e.g., $\lambda=0.1$ here, by decreasing the LR power $\alpha$, we can expect a growth of the correlation length because the critical point is effectively approached. 
This explains the slower decaying rate of $\mathcal{S}(r)$ for a smaller $\alpha$. 
Hence, the LR interaction does not significantly affect the AFM long-range order for $\alpha \textgreater 0$, which is why we do not focus on this region in this work.

However, motivated by Refs.~\cite{Vodola_2016, Koffel2012prl}, we are particularly interested in the gapped SPT regime (for $\lambda$ above the critical point). 
As shown in Fig.\ref{fig:correlation} (b) and (c), we found that for large $\alpha \gtrsim 1.0$, the SPT string order parameter exhibits a long-range order, while the connected spin-spin correlation displays an exponential decay at short distances and a power-law decay at longer distances.
Conversely, for sufficiently small $\alpha \lesssim 1.0$, both the connected spin-spin correlation function and the string order parameter decay algebraically at all distances; 
for this reason, we named this phase the algebraic SPT as mentioned in the main text. 
In addition, we examined the power exponent of the connected spin-spin correlation $|C_{\rm c}^{zz}(r)| \sim 1/r^{\gamma_{\rm s}}$ as a function of $\alpha$ along a fixed $\lambda$ line.
The result shown in Fig.~\ref{fig:correlation} (d) demonstrates that there is a clear crossover as $\alpha$ decreases below $\alpha_{\rm c} \approx 1.0$;
the estimated value of $\gamma_{\rm s}$ is approximately proportional to $\alpha$ when $\alpha > 1.0$, while it is roughly a constant when $\alpha < 1.0$\,.
It is noted that the change of the behavior of $C_{\rm c}^{zz}(r)$ is merely induced by the LR interaction. 
Considering that we are actually moving away from the critical point (line) as $\alpha$ decreases with $\lambda$ fixed, it is still possible that $\lambda$ can have an effect on the concrete values of $\gamma_{\rm s}$.
As exhibited in Fig.~\ref{fig:correlation} (d), the results of the power exponent $\gamma_{\rm s}$ along $\lambda = 0.55$ and $0.65$ lines have no obvious difference, which excludes this possibility directly. 


\subsection{Section VI: Edge mode splitting and the breakdown of the conformal symmetry at the topological nontrivial critical points}
\label{sec:SM6}

In this section, we provide further evidence to identify the topological properties of the LR quantum critical points in the phase diagram. 
In the short-range limit $\alpha \rightarrow \infty$, the model reverts to the nearest neighbor cluster Ising model, which exhibits a symmetry-enriched quantum critical point at $\lambda = 0.5$ with nontrivial topologically protected edge modes~\cite{verresen2021prx,wang2023stability}. 
In particular, the edge modes at the critical point display an unconventional algebraic finite-size splitting, scaling as $1/L^{14}$, which has attracted a great interest among theoretical physicists. 
According to Ref.~\cite{verresen2021prx}, this splitting can be revealed by adding a symmetry-preserving boundary perturbation, $ g (\sigma_{1}^{x} \sigma_{2}^{z} \sigma_{3}^{z} + \sigma_{L-2}^{z} \sigma_{L-1}^{z} \sigma_{L}^{x}) $, where $g$ is a small quantity. 
This perturbation can flip the boundary spins and couple the two edge modes, $\ket{\uparrow_{L}\downarrow_{R}}$ and $\ket{\downarrow_{L}\uparrow_{R}}$, leading to a finite-size splitting within the two-fold degenerate ground-state manifold. 
To say that the degeneracy of the edge modes is stable, it is necessary that the edge mode splitting is smaller than the bulk energy gap; in other words, the algebraic power needs to be larger than the dynamical exponent $z$.
For the short-range case, the power of the algebraic edge mode splitting is $14$ which is much larger than $z=1$, establishing the stability of the edge modes coexisting with the critical bulk fluctuations.

\begin{figure}[tb]
    \includegraphics[width=1.0\linewidth]{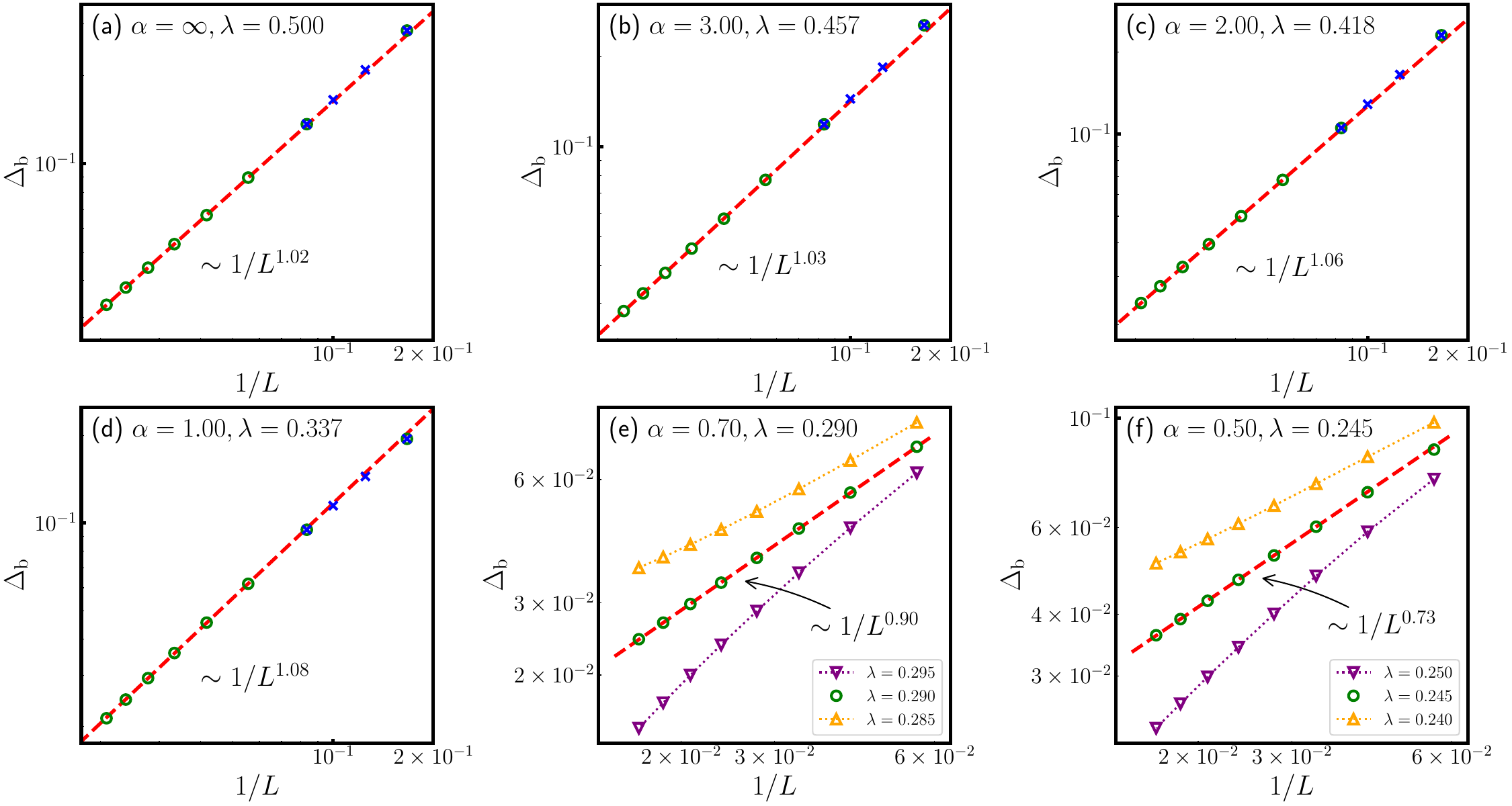}
    \caption{The finite-size scaling of the bulk energy gap $\Delta_{\rm b}$ at the critical points is depicted for typical values of $\alpha$: (a) $\alpha = \infty$, (b) $\alpha = 3.0$, (c) $\alpha = 2.0$, (d) $\alpha = 1.0$, (e) $\alpha = 0.7$, and (f) $\alpha = 0.5$\,. The critical points $\lambda_{\rm c}$ are determined by the peak position of $c_{\rm eff}$ for $\alpha \ge 1$; for $\alpha=0.7$ and $\alpha=0.5$, the critical point is determined by the observation of a linear behavior of $\Delta_{\rm b}$ in the log-log plot. The blue crosses in (a)-(d) are obtained from ED and the open markers are DMRG data. The red dashed lines are least-squares fittings according to $\Delta_{\rm b} \sim 1/L^{z}$\,.}
    \label{fig:bulkgap_cp}
\end{figure}

\begin{figure}[tb]
    \includegraphics[width=1.0\linewidth]{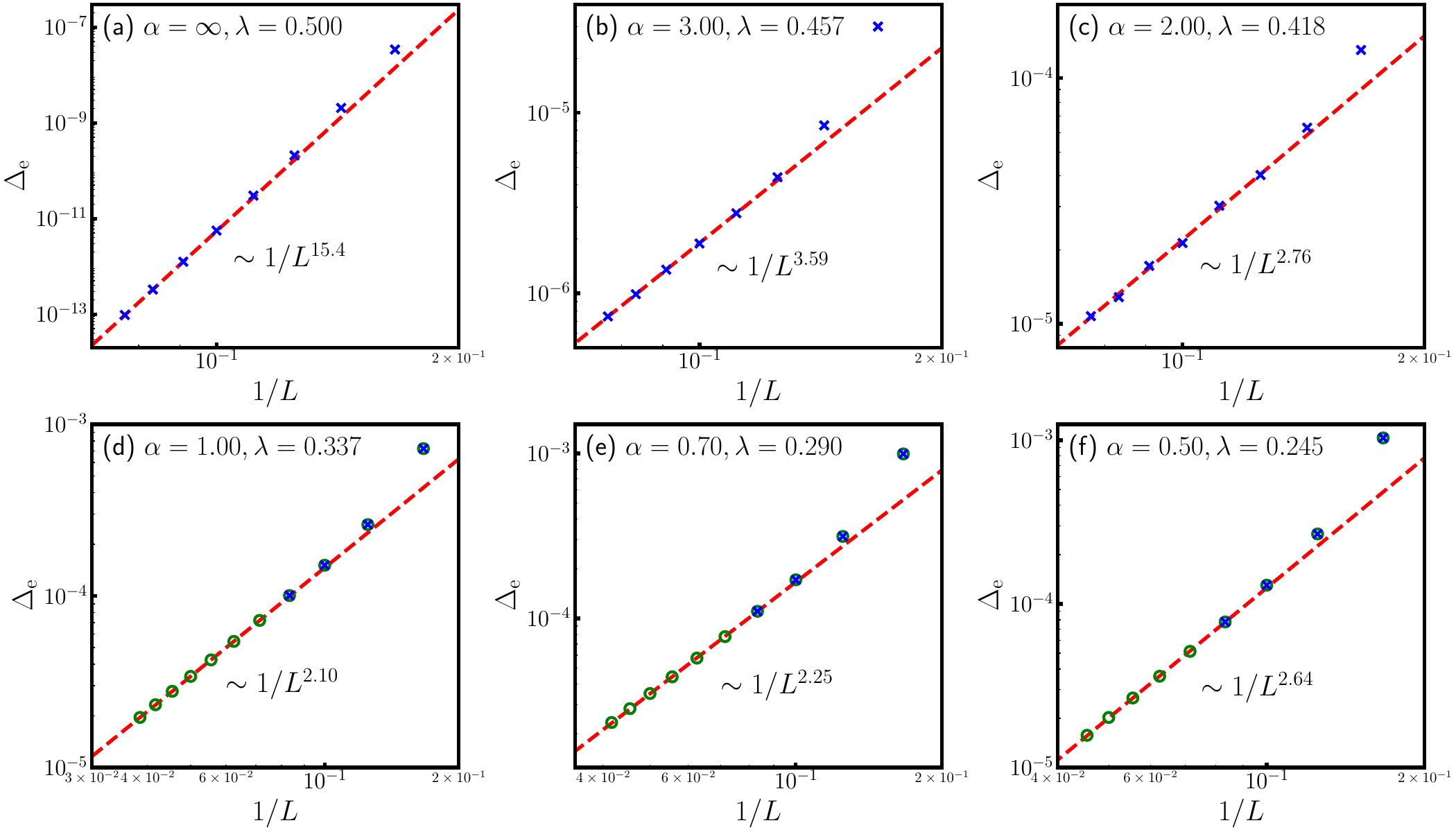}
    \caption{The finite-size scaling of the edge mode splitting $\Delta_{\rm e}$ by including a small boundary perturbation $0.1 (\sigma_{1}^{x} \sigma_{2}^{z} \sigma_{3}^{z} + \sigma_{L-2}^{z} \sigma_{L-1}^{z} \sigma_{L}^{x})$ at the critical points is depicted for typical values of  $\alpha$: (a) $\alpha = \infty$, (b) $\alpha = 3.0$, (c) $\alpha = 2.0$, (d) $\alpha = 1.0$, (e) $\alpha = 0.7$, and (f) $\alpha = 0.5$\,. The blue crosses are obtained from ED calculations and the green open circles are DMRG data. The red dashed lines are least-squares fittings according to $\Delta_{\rm e} = a/L^{b}$\,.}
    \label{fig:edgegap_cp}
\end{figure}

Now, a natural question to ask is whether the algebraically localized edge modes at the critical point can exist and be robust with the LR interactions? 
To answer this question, we first need to extract the dynamical exponent $z$, which enters the finite-size scaling of the critical bulk gap, $\Delta_{\rm b} \equiv (E_{3} - E_{1}) \sim 1/L^{z}$.
For a conformal invariant quantum critical point, we have $z = 1$.
As illustrated in Fig.~\ref{fig:bulkgap_cp}, we have performed a standard finite-size scaling analysis for the bulk gap $\Delta_{\rm b}$ at the critical points identified by the effective central charge $c_{\rm eff}$ (see Sec.~II for details).
It is found that the dynamical exponent $z$ remains to be $1.0$ within the numerical accuracy for large $\alpha \gtrsim 1.0$, suggesting that the conformal symmetry can be preserved when the power-law LR interaction is introduced.
The perfect power-law scaling observed in Fig.~\ref{fig:bulkgap_cp} (a)-(d) also verifies the high accuracy of the critical points located by the effective central charge $c_{\rm eff}$.
Conversely, for small $\alpha < 1.0$, we found that the critical point obtained from $c_{\rm eff}$ suffers from a small finite-size effect, and the scaling of the bulk gap deviates slightly from a power-law behavior. 
However, the peak position of $c_{\rm eff}$ still provides a good estimation of the critical point.
As shown in Fig.~\ref{fig:bulkgap_cp} (e)-(f), a perfect power-law scaling of $\Delta_{\rm b}$ can be found at $\lambda = 0.290$ for $\alpha = 0.70$ and $\lambda = 0.245$ for $\alpha = 0.50$, respectively. 
The result unequivocally demonstrates that the dynamical exponent $z$ is smaller than $1$ when $\alpha < 1$, implying a breakdown of the conformal symmetry.

To investigate the properties of the doubly degenerate edge modes at the critical points, we added the boundary perturbation mentioned above with a strength $g=0.1$ and computed the edge mode splitting as a function of $1/L$ for typical values of $\alpha$, as displayed in Fig.~\ref{fig:edgegap_cp}. 
The result shows that the edge mode splitting always exhibits an algebraic decay regardless of the strength of the LR interactions. 
More importantly, by comparing with the dynamical exponent $z$ obtained in Fig.~\ref{fig:bulkgap_cp}, we can find that the power of the algebraic edge mode splitting is always larger than the corresponding dynamical exponent $z$ at least for the $\alpha \gtrsim 0.5$ considered here. The calculation for $\alpha < 0.5$ is more challenging because an accurate determination of the critical point requires to simulate systems of larger size.
It is noted that the power exponent obtained in Fig.~\ref{fig:edgegap_cp} can be effected by finite-size effects. For example, the power exponent in the short-range limit is predicted to be $14$ in the thermodynamic limit, and there is a finite-size correction $\sim 1/L^{16}$~\cite{verresen2021prx} leading to a slightly larger estimation given in Fig.~\ref{fig:edgegap_cp} (a) for example.
As the concrete form of the finite-size correction is unknown for finite values of $\alpha$, we can not perform a systematic extrapolation to the $L \to \infty$ limit. 
However, we can expect that the finite-size effect can bring about $(15.4-14)/14 = 10\%$ estimation error for finite $\alpha$ (especially for $\alpha \lesssim 1$ because larger system sizes are used in the least-squares fittings).
In summary, the result indicates that even with LR interactions, the algebraically localized edge modes persist at the critical point. 
This is a typical feature of topologically nontrivial critical points and provides further evidence to confirm the conclusion  we drawn in the main text.

\subsection{Section VII: Details of analytical calculation for $\alpha=0$}
\label{sec:SM7}

\paragraph{Ground state degeneracy in the SPT phase and critical point} 
In this section, we explain the degeneracy found in the entanglement spectrum. The degeneracy in entanglement spectrum of periodic chain corresponds to the degeneracy in the energy spectrum of open chain. In the following, we consider an open finite chain with size $L$. Deep in the SPT phase, $\lambda=1$, the local operators $\sigma_1^z$, $\sigma_L^z$ commute with the Hamiltonian results in 4-fold ground state degeneracy,
\begin{equation}
    \ket{\uparrow_L\uparrow_R} \pm \ket{\downarrow_L\downarrow_R}, \quad \ket{\uparrow_L\downarrow_R} \pm \ket{\downarrow_L\uparrow_R},
\end{equation}
These 4 degenerate states correspond to the 4-fold degeneracy found in the entanglement spectrum. 

At the critical point, the long-range Ising interaction generates the AFM interaction between the left boundary and right boundary. It favors 
\begin{equation}\label{eq:bdymodes}
    \ket{\uparrow_L\downarrow_R} \pm \ket{\downarrow_L\uparrow_R}
\end{equation}
and lifts the other two states with a $\sim 1/L^z$ gap, where $z$ is the dynamical exponent. Therefore, the critical point only hosts 2-fold degenerate ground states.

We note that the the gapped phase under the $\alpha\rightarrow 0$ limit also hosts 2-fold degenerate ground states. Since the all-to-all Ising interaction contains the coupling between the leftmost and rightmost boundary spins and favors \eqref{eq:bdymodes}, while lifts the other two states with an order $1$ gap.

\paragraph{Properties in $\alpha \rightarrow 0$ limit}
In this paragraph, we explain the emergence of a new \emph{gapped} SPT phase with two-fold degenerate edge modes in the $\alpha\rightarrow 0$ limit. In the $\alpha \rightarrow 0$ limit, the first term in the Hamiltonian \eqref{E1} is all-to-all Ising interaction, 
\begin{equation}\label{Hinfty}
    H_\infty = \frac{(1-\lambda)}{N}\sum_{i<j} \sigma_i^z \sigma_j^z +\lambda \sum_{j} \sigma^z_j \sigma^x_{j+1} \sigma^z_{j+2}
\end{equation}
where $N$ is the total sites. The spectrum of the model is unchanged when applying unitary transformations. This model can be further mapped to infinite-range transverse field Ising model by conjugating the Hamiltonian with $U=\prod_i CZ_{i,i+1}$, where,
\begin{equation}
    CZ_{i,i+1} = \begin{pmatrix}
        1&0&0&0\\ 0&1&0&0  \\ 0&0&1&0  \\0&0&0&-1
    \end{pmatrix}
\end{equation}
The transformed Hamiltonian is,
\begin{align}
    H'_\infty &= U^\dagger H_\infty U = \frac{(1-\lambda)}{N}\sum_{i<j} \sigma_i^z \sigma_j^z +\lambda \sum_{j} \sigma^x_{j} \\
    &= \frac{(1-\lambda)}{2N}(\sigma^z_{tot})^2   +\lambda  \sigma^x_{tot} -\frac{(1-\lambda)}{8} \label{hpinfty}
\end{align}
where $\sigma^{a}_{tot} = \sum_i \sigma_i^a$. The algebraic SPT phase is mapped to the symmetric paramagnetic phase in the infinite-range transverse field Ising model. The infinite-range transverse field Ising model is effectively a 0+1d quantum mechanics model of spin $\frac{N}{2}$. The total spin $S(S+1)$ is a good quantum number, and $0\le S\le \frac{N}{2}$. 

In the paramagnetic phase $\lambda>\lambda_c$, we use the Holstein-Primakoff transformation,
\begin{equation}
    \sigma_{tot}^x = S-a^\dagger a,\quad \sigma_{tot}^z -i \sigma_{tot}^y  = \sqrt{2S-a^\dagger a} a,\quad \sigma_{tot}^z +i \sigma_{tot}^y  = a^\dagger\sqrt{2S-a^\dagger a}.
\end{equation}
where $[a,a^\dagger] = 1$. Since the low energy properties of the system are dominated by $S$ closed to $\frac{N}{2}$, we can approximate the Holstein-Primakoff transformation to be,
\begin{equation}
    \sigma_{tot}^x = S-a^\dagger a,\quad  \sigma_{tot}^z  = \sqrt{\frac{S}{2}}(a+a^\dagger).
\end{equation}
Then the Hamiltonian \eqref{hpinfty} becomes,
\begin{equation}
    H'_\infty = \frac{(1-\lambda)}{4N}(a+a^\dagger)^2   +\lambda (S-a^\dagger a) -\frac{(1-\lambda)}{8}
\end{equation}
One can easily find the spectrum is given by,
\begin{equation}
    E_{S,n} = -(-\lambda)(S+\frac{1}{2})+(n+\frac{1}{2})(-\lambda)\sqrt{1-\frac{(1-\lambda)S}{\lambda N}} -\frac{(1-\lambda)}{8}
\end{equation}
The gap between ground state and the first excited is a constant which indicates the $z=0$.

In the $\alpha \rightarrow 0$ limit, all the two point correlation functions are equal $\langle \sigma_i^z \sigma_{j}^z \rangle = \text{const}.$, while the constant varies with $\lambda$. This is related to the fact that system has a large site permutation symmetry, and it is effectively a $0+1d$ system.

\subsection{Section VIII: Experimental implementation}
\label{sec:SM8}

In this section, we discuss the potential experimental realization of our model. 
We note that Refs~\cite{chen2023scipost,shen2023observation} propose that generic LR spin models can be implemented in a quantum circuit. 
Specifically, we can transform the 1D spin chain into an IBM quantum qubit device suitable for quantum circuits. 
To investigate the ground state of our model, we employ the Quantum Imaginary Time Evolution (QITE) method~\cite{kamakari2022prxq}. 
In our circuit design, the non-unitary process of QITE is effectively realized by coupling the main circuit, constructed from the sites, with a single ancilla qubit.

First, we prepare the non-unitary dynamics $U_{N}=e^{-\beta H}$ of our LR spin system with $N$ qubits. 
This step is programmable and can be achieved through the tensor network method. 
Thus, the LR Ising coupling and cluster interaction are numerically incorporated into our target non-unitary operation $U_{N}$. 
Subsequently, this non-unitary $U_{N}$ is numerically embedded in a larger unitary process $U_{N+1}$ of $(N+1)$ qubits~\cite{shen2023observation}. 
It's worth noting that there's no need to physically implement the LR Ising coupling and cluster interaction terms in our model;
what needs to be realized is this extended unitary $U_{N+1}$. 
This setup enables us to perform final post-selection on the ancilla qubit, leading to effective QITE on the physical system. 
For practical implementation on a quantum computer, i.e., realizing the unitary $U_{N+1}$, we can utilize variational circuit optimization. 
Here, we prepare a parameterized circuit $V$ with relatively short depth and then employ variational optimization to approximate our target circuit: $U_{N+1} \approx V$.

\end{document}